\documentclass{article}

\DeclareUnicodeCharacter{2003}{\'k}

\usepackage{PRIMEarxiv}

\usepackage[utf8]{inputenc} 
\usepackage[T1]{fontenc}    
\usepackage{hyperref}       
\usepackage{url}            
\usepackage{booktabs}       
\usepackage{amsfonts}       
\usepackage{nicefrac}       
\usepackage{microtype}      
\usepackage{lipsum}
\usepackage{graphicx}
\usepackage[ruled,vlined]{algorithm2e}
\usepackage{amsmath,amsfonts,amssymb}

\title{Temporal Super-Resolution using Multi-Channel Illumination Source
}

\author{
  Khen Cohen \\
  Electrical Engineering \\
  Tel Aviv University \\
  Tel Aviv\\
   \And
  Dan Raviv \\
  Electrical Engineering \\
  Tel Aviv University \\
  Tel Aviv\\
   \And
  David Mendlovic \\
  Electrical Engineering \\
  Tel Aviv University \\
  Tel Aviv\\
}

\begin{document}
\maketitle

\begin{abstract}
While sensing in high temporal resolution is necessary for wide range of application, it is still limited nowadays due to cameras sampling rate. In this work we try to increase the temporal resolution beyond the Nyquist frequency, which is limited by the sampling rate of the sensor. This work establishes a novel approach for Temporal-Super-Resolution that uses the object reflecting properties from an active illumination source to go beyond this limit. Following theoretical derivation, we demonstrate how we can increase the temporal spectral detected range by a factor of 6 and possibly even more. Our method is supported by  simulations and experiments and we demonstrate as an application, how we use our method to improve in about factor two the accuracy of object motion estimation.
\end{abstract}


\section{Introduction}
Resolution in a digital signal refers to its frequency content: high-resolution (HR) signals are band-limited to a more extensive frequency range than low-resolution (LR) signals. The resolution is mostly limited by two factors: the physical device limitation and the sampling rate. For example, digital image resolution is limited by the imaging device's optics (the diffraction limit) and the sensor's pixel density (the sampling rate).

Super-resolution is a broad research area, which uses sophisticated ways to overcome these limits. The ability to exceed the system resolution limits always has something to do with some prior knowledge about the scene or about the system \cite{SR_information_capacity} \cite{sr_david_sb}. 
Image Super-Resolution (SR) techniques can be divided into two main approaches:

Optical SR - Utilizing the optical property of the light to transcend over the diffraction limit. There are mainly thee approaches in that field,
the first is multiplexing spatial-frequency bands \cite{sr_moire}, which uses the fact that low-frequency moire fringes are formed when the scene is multiplexed with a periodic pattern (structured illumination).
The second is by acquiring multiple parameters about the scene and merge them, for example, detect the scene polarization \cite{sr_vs_polarization}.
The third method is the probing near-field electromagnetic disturbance method. It is a modern approach that uses an unconventional imaginary optical system and tries to detect tiny disturbances in the electromagnetic waves. For example, using evanescent wave \cite{SR_probing}.
Each one of these super-resolution methods sacrifice another domain \cite{sr_david_sb}, \cite{sr_dof_of_img} \cite{SR_information_capacity}. For example on account of the time \cite{sr_vs_time}, wavelength \cite{sr_vs_wavelength} \cite{sr_vs_wavelength2} or field of view \cite{sr_vs_fov}.

The other type of Image super-resolution is called Geometrical SR. These methods focus on the sensor pixels density limit. It includes mainly algorithmic solutions such as frame de-blurring and localization estimator \cite{sr_geometric}. Nowadays, Deep Learning methods have presented an excellent performance on SR tasks, as shown in \cite{sr_geometric_dl}.

Temporal Super-Resolution (TSR) - While Spatial Super-Resolution has been widely researched for decades, TSR has not been extensively investigated until recently. In general, TSR can be divided into three main approaches: 
A combination of Cameras - This method exploits the fact that different cameras, with some temporal overlap, can provide complementary information to increase the temporal resolution.
Temporal Coding - method uses temporal pre-known pattern as a coding technique for the detected signal. Optical Coding extract temporal illumination patterns \cite{optical_tsr_illumination}  or temporarily coded aperture \cite{optical_tsr} and Sensor coding uses a temporary change of the sensor read manner \cite{tsr_sensor} or using flattered shutter \cite{tsr_debluring_shutter} to de-blur the images.
Software Interpolation - uses algorithms (mainly Deep Learning techniques) to generate temporal interpolation of the signal. Part of the methods are Optical Flow based \cite{tsr_of}, some are phase based \cite{tsr_phase_shift} and Kernels based methods \cite{tsr_kernel_based} \cite{tsr_kernel_based2}.

Software only approaches a straightforward solution in terms of system complexity, and these methods demonstrate good performances \cite{2020tsr}. However, their ability to interpolate in time is limited since the Deep Learning models heavily rely on past examples and training. In contrast, TSR supported by hardware (optics or sensor) has the potential to raise the temporal sampling frequency with a much higher rate and reliability. The price, however, is the complexity of the system.
Despite the reviewed research in this field, no practical solution with a high up-sampling factor, high reliability, and low system complexity has been shown yet.

In this work, we present a novel approach for TSR, using the object's optical reflection properties such as its surface polarity reflection or the spectral reflectively (the object's color). Our proposed system is composed of a standard camera with a high-frequency illumination source. We model the camera image sensor operation method, formulate our problem as an optimization problem and provide a comprehensive solution for flicker with RGB colors. Our analysis and results are supported by theoretical derivations, simulation, and experimental results.
Apart from other works in this field, our method demonstrate high reliability of spectral reconstruction with no significant hardware complexity penalty and it performs actual Temporal Super-Resolution method, which means our method can deal with wide range of cases, and not only for blurred images. Moreover, our method can be used in real time due to its simple solution form.
\newline
The main contributions of this work is as follows:
\begin{enumerate}
  \item Demonstration of a novel approach for optical coding to achieve high temporal frequencies with fixed sensor sampling rate.
  \item Develop a substantial theoretical background to increase temporal resolution from sub-samples.
  \item Provided Anti-Aliasing algorithm to improve the system performance in wide range of frequencies.
  \item Real-time reliable technique to perform Temporal Super-Resolution
\end{enumerate}

\section{Theoretical Background}

\subsection{Temporal Model for an Image Sensor}
We denote a general signal as $I(x,y,t)$, captured by the image sensor.
We formulate the image sensor operation as temporal distortion, which is assumed to be linear time-invariant (LTI), followed by sampling. The distortion is represented by the transfer function $h(t)$, and sampling in time at a frequency of one over the exposure time $f_s=\frac{1}{T}$.
The sampled signal, therefore, is given by:
\begin{equation}
u[n] \equiv u(t = nT) = \left[ \sum_{n=-\infty}^{\infty} \delta \left(t-Tn \right)  \right]
    \left[ f(t) * h(t) \right] \sum_{n=-\infty}^{\infty} \delta \left(t-Tn \right)
    \left[ f(t) * h(t) \right]_{t=nT}
\end{equation}
In order to fully reconstruct the $u(t)$ signal, two conditions have to be fulfilled: First, the distortion of the signal is not too severe, and the sampling rate is at least twice as high as the maximum spectral content of $u(t)$, according to Nyquist theorem of sampling.
Raising the sampling rate should improve the signal reconstruction. Even though, according to the second condition, it is not entirely clear because raising the sampling rate means shortening the integration time, which is accompanied by Signal To Noise (SNR) degradation.

\begin{figure}[hbt!]
    \centering
    \includegraphics[trim={0cm 0cm 0cm 0cm},width=11cm, height=8.5cm]{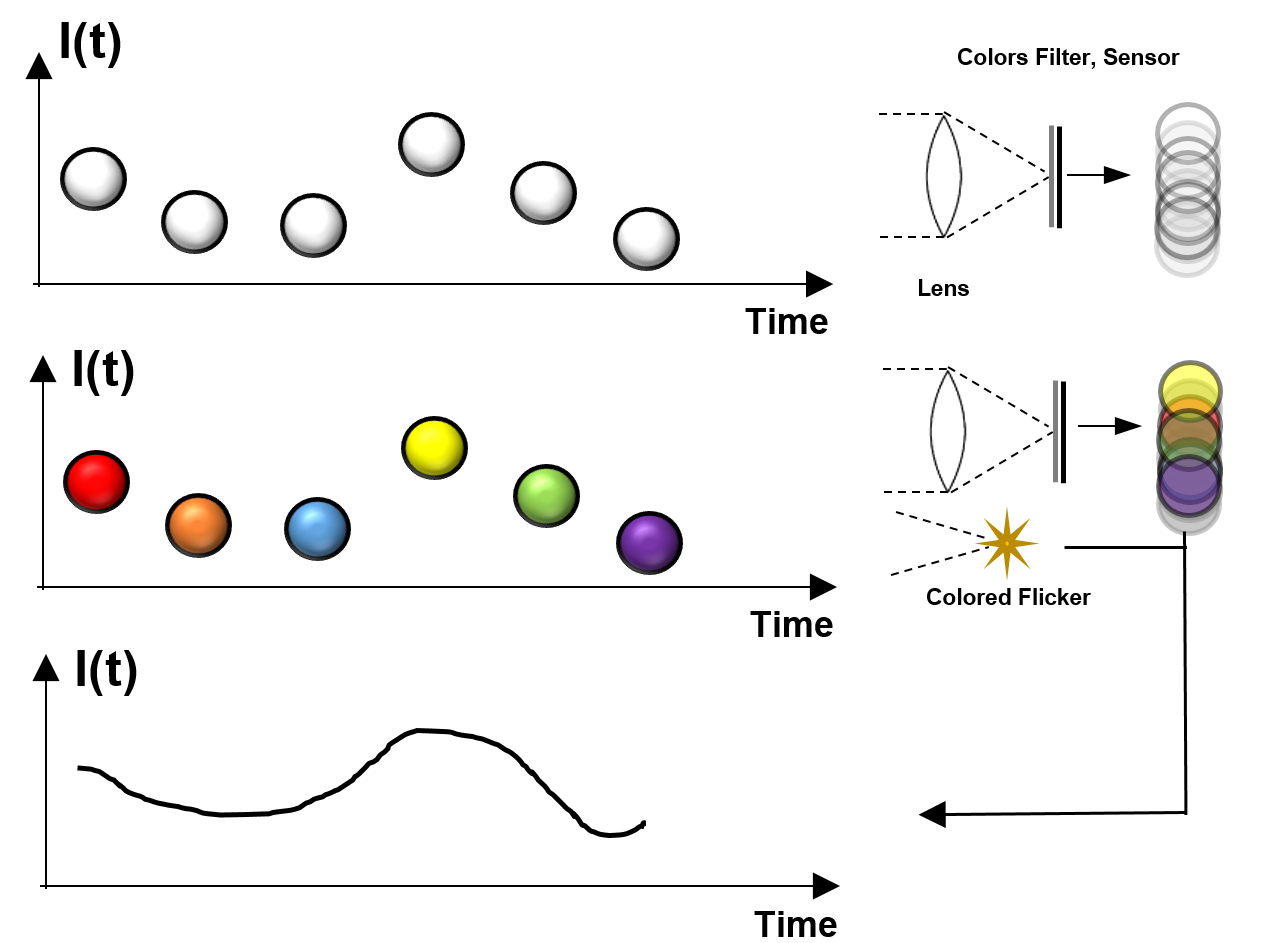}
    \caption{Schematic Diagram of our method. An object moves and changes its intensity value under different illumination conditions: Top - arbitrary environment illumination. In this case, there is no obvious way to reconstruct the object intensity value in time since the sensor integrates over all the light from the scene. Middle - under colored flicker illumination with our prior knowledge about the flicker pattern, we can recover the object intensity value in high quality of certainty. Among all the possible temporal profiles, we choose the most reasonable one in the sense of minimum energy.}
    \label{fig:TSR_diagram}
\end{figure}

\subsection{Multi-channel approach, assumption and solution}
We define a set of channels as a set of independent optical properties of the light. For example, X-polarization and Y-polarization are two channels, or several different wavelengths are different channels. We assume linear optics, meaning that the reflected light from an object does not transform between channels. We define the channel $m$ as follows:
\begin{equation}
    C^{m} = \int_{0}^{T} \int_{-\infty}^{\infty} c^{m}(\lambda,t) Q^{m}(\lambda) R(\lambda,t) d\lambda dt
\end{equation}
While $c(\lambda,t)$ is the illumination mask generated by a light source (changes in time), $R(\lambda,t)$ is the reflectively properties of the object (change in time) and $Q(\lambda)$ is the image sensor filter for a specific spectral range, $\lambda$ is the wavelength and $T$ is the integration time of the sensor (exposure time).

We assume that for a given $m$, there is a spectral match between the flicker light source and the sensor filter. In addition, we assume $c^m(\lambda,t)$ is a product of temporal-dependant function and spectral-dependent function:
\begin{equation}
    c^m(\lambda,t)Q^m(\lambda) =  c^m(t)\tilde{Q}^m(\lambda)
\end{equation}
Moreover, we assume that there is high similarity between the different channels. So that for each time $t$ the relation between the collected light at each channel is equal up to a constant scale $\gamma$:
\begin{equation}
    \int_{-\infty}^{\infty} \tilde{Q}^{m}(\lambda) R(\lambda,t) d\lambda \approx \gamma^{m,k} \int_{-\infty}^{\infty} \tilde{Q}^{k}(\lambda) R(\lambda,t) d\lambda
\end{equation}
We focus now on the case when the flicker changes in time in a discrete manner between two modes, off and on. Therefore we get:
\begin{equation}\label{equation:channel_definition}
    C^{m} = \sum_{n=1}^{N} c^{m}_{n} \int_{-\infty}^{\infty} \tilde{Q}^{m}(\lambda) R(\lambda,t) d\lambda \approx \sum_{n=1}^{N} c^{m}_{n} i_{n}
\end{equation}
While N is the up-sampling factor and $i_{n}$ represents the average value of the image at sub-time-step $n$.

\subsection{Definitions}
 our analysis, we denote $N$ as the up-sample factor of the sampling rate, meaning that we increase the maximum detected spectrum from $\frac{1}{2T}$ to $\frac{N}{2T}$, and $M$ is the number of independent channels we use.
We assume that in any sub-interval of time $\frac{T}{N}$ is approximately constant, so we can define for each exposure time the intensity vector of size N, $\vec{I}$. We further define the vector $\vec{C}$, of size M, to represent the captured value in each of the channels for a single exposure time.
We define M vectors ($m$ is between 1 to M) the vectors $\vec{c}^{m}$, in size N, to represent each of the channels' code patterns. In our analysis, we focus on the cases where the vectors $\vec{c}^{m}$ have binary values, 0 or 1, when the flicker of channel $m$ is on or off, respectively.
In addition, we define a vector in size M to represent the Lagrange multiplier of each of the channels $\vec{\lambda}$
Finally, we define the matrices $\textbf{S}$, $\textbf{M}$, which will be used in our derivations:
\begin{equation}
    \textbf{ S }_{(N,M)} = \begin{pmatrix} 
     \vec{c}^1 & \dots & \vec{c}^{M} \\ 
    \end{pmatrix},
    \hspace{0.5cm}
    \vec{ C }_{(M,1)} =
    \begin{pmatrix} 
        C^{1} \\ 
        \vdots \\ 
        C^{M} 
    \end{pmatrix}
    \hspace{0.5cm}
    \vec{ \lambda }_{(M,1)} =
    \begin{pmatrix} 
        \lambda^{1} \\ 
        \vdots \\ 
        \lambda^{M} 
    \end{pmatrix}
\end{equation}
\begin{equation}
    \textbf{ M }_{(N,N)} = \begin{pmatrix} 
    4 & -2 & 0  &\dots & 0\\ 
    -2 & 4 & -2 &\dots & 0\\ 
    0 & -2 & 4 & \dots & 0\\ 
    & & \vdots & & & \\
    0 & \dots & -2 & 4 & -2 \\ 
    0 & \dots & 0 & -2 & 4 \\ 
    \end{pmatrix},
    \hspace{0.1cm}
    \vec{ I }_{(N,1)} = \begin{pmatrix} 
    i_1 \\
    \vdots \\
    i_N 
    \end{pmatrix}
\end{equation}

In our work, we will present an extension model to utilize the spatial correlation between adjacent pixels, for those parts, we redefine the vectors and matrices as follows:
\begin{equation}
    \vec{ C }_{(M*5,1)} =
    \begin{pmatrix} 
        C^{1}_{1} \\ 
        \vdots \\ 
        C^{M}_{1} \\
        \vdots \\
        C^{1}_{5} \\ 
        \vdots \\ 
        C^{M}_{5} \\
    \end{pmatrix}
    \hspace{0.5cm}
    \vec{ \lambda }_{(M*5,1)} =
    \begin{pmatrix} 
        \lambda^{1}_{1} \\ 
        \vdots \\ 
        \lambda^{M}_{1} \\
        \vdots \\
        \lambda^{1}_{b,5} \\ 
        \vdots \\ 
        \lambda^{M}_{r,5} \\
    \end{pmatrix}
\end{equation}
S matrix ($5Nx5M$) is changed to be a block diagonal matrix while each block correspond to different pixel in the domain P.
M matrix ($5Nx5N$) has changed to be:
\begin{equation}
    \textbf{ M }_{i,j} = 
    \begin{cases}
        2w^s+2w^t & \text{if} \ i = j  \\
        -2w^t & \text{if} \ |i - j| = 1  \\
        -2w^s& \text{if} \ |i - j| mod 5 = 0  \\
        0 & \textbf{else} \ \\
    \end{cases}
\end{equation}

We define the square-wave signal as follows:
\begin{equation}
    SW_f(t) \equiv sgn\left[sin(2\pi f t)\right]
    \hspace{0.5cm}
    sgn(x) \equiv \left\{  \begin{array}{ll}
          1 & x > 0 \\
          -1 & x < 0 \\
          0 & x = 0 \\
    \end{array}\right.
\end{equation}

\subsection{Signals Similarity Metrics}
To evaluate the method's accuracy, we shall compare our result with the actual signal. Since we deal with function with finite energy (in $L^2$), the inner product is defined as:
\begin{equation}
    <f_1(t),f_2(t)> \equiv \int_{-\infty}^{\infty} f_1(t)f_2(t)dt
\end{equation}

We introduce two error metrics we found the most reasonable to examine:

\textbf{Euclidean distance (L2)} - 
\begin{equation}
    L_2(f_1(t),f_2(t)) \equiv \sqrt{||f_1(t)-f_2(t)||_2^2}
\end{equation}

\textbf{The Cosine similarity} - 
\begin{equation}
    Error(f_1(t),f_2(t)) \equiv \cos^{-1}\left( \frac{< f_1(t), f_2(t) >}{\sqrt{ ||f_1(t)||^2 ||f_2(t)||^2}} \right) 
\end{equation}
The first metric was chosen because it is a prevalent and intuitive way to compare signals.
The second metric was chosen to avoid bias due to the signals' absolute magnitude and focus only on their shape relations; it was used in the experimental part.

\section{Method}


From equation \ref{equation:channel_definition} one can notice that extracting the values of $i_{n}$ is equivalent to up-sample in factor N at the time domain. The problem is that for $M$ channels, this equation can be solved uniquely only for an up-sample factor of $N=M$.
In practice, we have low number of channels and we want to get high rate of temporal-super-resolution. For that, we need to use some prior knowledge about the scene dynamics. We choose to assume scene smoothness in time and to formulate this we define the following cost function:
\begin{equation} \label{equation:cost_function}
    \mathcal{L} = \sum_{n=1}^{N-1} \left( i_n-i_{n+1} \right)^2 + \sum_{m=1}^{M} \lambda^m \left( C^m - \sum_{n=1}^{N-1} i_n c^{m}_{n} \right)
\end{equation}
While $\lambda^{m}$ are Lagrange multipliers.
Find the solution of \ref{equation:cost_function} means that, from the infinite number of solution to equation \ref{equation:channel_definition}, we would like to choose the one most smooth solution among all to be the estimator for the actual signal.
The solution is given by the following equation (for the complete derivation, see appendix):
\begin{equation} \label{equation:optimization_solution}
    \boxed{ \vec{I} = \textbf{M}^{-1} \textbf{S} \left( \textbf{S}^\intercal \textbf{M}^{-1} \textbf{S} \right)^{-1} \cdot\vec{C} }
\end{equation}

\subsection{Spatial Regularization}
\begin{figure} [h]
    \centering
    \includegraphics [width=0.2\linewidth] {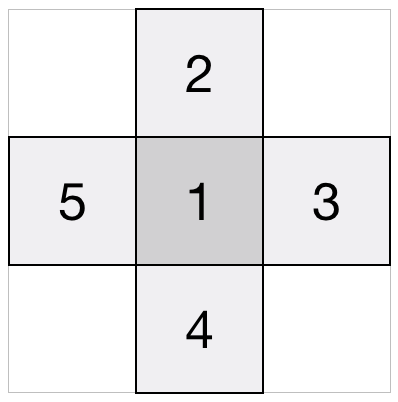}
    \caption{Pixels neighbors order as we used in our method. P domain includes pixels 2,3,4 and 5}
    \label{fig:pixels_neighbors}
\end{figure}

The absence of any spatial correlation between adjacent pixel might yield some artifacts in the image. To avoid this, we define for each pixel a domain $P$ which includes the pixel with its 4 closest neighbors (see figure \ref{fig:pixels_neighbors}), and we modify the cost function as follows:
\begin{gather*}
    \mathcal{L} = \sum_{n=1}^{N-1} \sum_{x,y \in P} w^{t}_{x,y} \left( i_{x,y,n}-i_{x,y,n+1} \right)^2 + 
    w^{s}_{x,y} \left( i_{x,y,n}-i_{x,y+1,n} \right)^2 + \\
    w^{s}_{x,y} \left( i_{x,y,n}-i_{x+1,y,n} \right)^2 +  \sum_{m=1}^{M} \sum_{x,y \in P} \lambda_{m,x,y} \left( C_{m,x,y} - \sum_{n=1}^{N} i_{x,y,n} c_{x,y,m,n} \right)
\end{gather*}
The vectors changes to be column stack vectors of the different pixels and the matrices are expanded to block-matrices as explained in the Appendix. $w$ factors are weight factors that determine the ratio between the spatial and temporal regularization.

\subsection{Colored Light Source}We focus on a particular case of flicker in colors Red, Blue, and Green. This case is the most common and can be used on any colored camera. We denote the flickers vectors as $\vec{c}^{1}=\vec{r}$,  $\vec{c}^{2}=\vec{g}$, $\vec{c}^{3}=\vec{b}$ and the channels vector is $\vec{C}^\intercal = (R, G, B)$, while $R$, $G$ and $B$ are the color digital value as it captured in the image for each color.

The spectral matching assumption as presented is fulfilled because the light source spectrum is similar to the sensor filter spectrum.
The high similarity assumption, as presented previously, gets now the following interpretation (in digital values):
\begin{equation}
    u(t) \approx \gamma^r u_r(t) \approx \gamma^g u_g(t) \approx \gamma^b u_b(t)
\end{equation}
While $u(t)$ is the true signal and $u_r(t)$, $u_g(t)$, $u_b(t)$ are the signals as it captured in color red, green and blue respectively. The $\gamma$ factors are related to the object color, and they can be derived from a single image of the scene (with no flicker).

A binary flicker pattern of red, green, and blue for each camera's exposure time illuminates the scene. The total accumulated result is used to extract the value of the actual signal (see diagram \ref{fig:TSR_diagram})

\subsection{The Scanning Mode}
Since $N$ represents the up-sample factor, the smaller the $N$, the more accurate the result should be (for $N=3$, the result is even unique), but no information about higher frequencies is collected. In contrast, the high $N$ factor can detect high-frequency content but is less reliable.
Therefore, we propose a technique that applies several $N$ factors, each at a separate temporal window. At the same time, we define a temporal window as a period in which the method works at a constant $N$ factor.

\textbf{Construction of the signal} - The construction of the signal is done in the spectral domain. However, collecting all the contributions from the different temporal windows is not straightforward. There could be many approaches to combination strategy. We chose the following: Each spectral interval of the united signal is given by averaging over all the temporal windows' contributions with the minimum $N$ factor that detects this spectral interval. For example, given a camera with a frame-per-second rate of $f_s$, if we apply a scanning method with the sequence: $N = 3, 4, 5, 5$, The low spectral domain (up to $3\frac{f_s}{2}$) is equal to the spectral content of the first temporal window, the mid-spectral-domain (from $3\frac{f_s}{2}$ to $2f_s$) is equal the spectral content of the second temporal window and the high spectral content (from $2f_s$ to $5\frac{f_s}{2}$) is equal to the average between the third and the fourth temporal windows.

One assumption that underlies this method's basis is that the spectral content of the scene does not change much between temporal windows (invariant signal for short time).
According to that, choosing as short temporal windows as possible is preferred, yet, too short temporal windows may not provide enough accurate results for the spectrum.

\textbf{Anti-Aliasing algorithm} - Each spectral contribution for each temporal window contains the influence of aliased signals from higher frequencies. Even though, after we construct the whole signal, we can use mutual information from the different spectral domains in order to attenuate and even eliminate the aliasing \ref{alg:AA}.

\begin{algorithm}[H]
\SetAlgoLined
\KwResult{ $I$ - The Signal with no Aliasing }
 initialization\;
 \For{N in Inverse Sorted L}{
    $f_{min}$ = $(N-1)\cdot\frac{FPS}{2}$\; 
    $f_{max}$ = $N\cdot\frac{FPS}{2}$ \;
    $A$ = BPF($f_{min}$, $f_{max}$)\;
    $A$ = Rotate($I$, $f_{min}$)\;
    $I$ $=$ $I-A$
 }
 \caption{Anti-Aliasing algorithm}\label{alg:AA}
\end{algorithm}

While BPF is an ideal Band-Pass-Filter, Rotate is a function that rotates the signal's spectrum relative to a specific frequency.
The algorithm uses the fact that every temporal window with a specific $N$ is aliased mainly by the spectral components from the $N+1$ temporal window's components. In this way, we use the spectrum that the temporal window had recovered with the up-sampling factor of $N+1$ and subtract its aliasing contribution from the spectral range recovered by the temporal window using the $N$ up-sampling factor.

\subsection{Performance analysis and Signal to Noise Ratio (SNR)}
As presented in previous sections, typically, increasing the FPS goes with decreasing the exposure time. The SNR of the signal grows linearly with the exposure time. Hence, reducing the exposure time should decrease the SNR. However, the SNR grows like a square root in the illumination intensity (or the number of photons) \cite{SNR_model}. Since our method uses an active illumination source, it compensates for the SNR decrease and improves image quality. We analyze the signal and the noise separately:

\textbf{The Signal} - 
We split the exposure time to N disjoints time steps according to the derivation. At the same time, For each of these steps, we assumed that the captured light was generated by the light source and then was reflected from the object (up to a color reflectively factor), in addition to the environment's existing background illumination. In that case, we can not assume that the captured signal has been taken only at the time steps when the flicker was applied. Therefore, a more precise derivation should be presented (we denote b = blue, g = green, and r = red, and $t$ is the exposure time of the camera):
\begin{align*}
    S_{TSR} = & \sum_{n=1}^{N}
    \frac{t}{N} \left[ 
    \gamma_b \left(i^{b\ env}_{n}+i^{b\ flicker}_{n}\right) + \gamma_g \left(i^{g\ env}_{n}+i^{g\ flicker}_{n}\right) +
    \gamma_r \left(i^{r\ env}_{n}+i^{r\ flicker}_{n}\right)
    \right]
\end{align*}

This equation says that for each frame, the signal comprises a summation of the intensity from the environment and the flicker over N time steps.
We assume that the flickering intensity is equal for different colors, and we denote it as $i^{flicker}$. We assume that the environment light is a constant white light with approximately equal intensity for the entire spectral range, denoting it as $i^{env}$. In addition, we denote the ratio $\alpha \equiv \frac{i^{flicker}}{i^{env}}$ so we can say that:
\begin{align}
    S_{TSR} \approx \left(\alpha^{-1}+1\right) i^{flicker}
     \sum_{n=1}^{N} \frac{t}{N} \left( \gamma_b b_n + \gamma_g g_n + \gamma_r r_n \right)
     \geq \left(\frac{1}{\alpha}+1\right) i^{flicker} t \min{ \{\gamma_b, \gamma_g, \gamma_r \} } \nonumber
\end{align}
We used the fact that we do not allow a time step without a flicker at all.
The Signal ratio between the signal with and without the flicker ($S_T$ represents the SNR without the flicker):
\begin{align}
    \frac{S_{TSR}}{S_T} = \frac{ \left(\alpha^{-1}+1\right) i^{flicker} \sum_{n=1}^{N} \frac{t}{N} \left( \gamma_b b_n + \gamma_g g_n + \gamma_r r_n \right) } {t i^{env}}
    \geq
     \left(1+\alpha\right) \min{ \{\gamma_b, \gamma_g, \gamma_r \} } \nonumber
\end{align}

\textbf{The Noise} - 
Say we deal with a temporal signal and separate its exposure time into several temporal steps. Then, we adopt the noise model as presented in \cite{SNR_model}. Note that the sensor still works on the same original exposure time, but we assume that the dominant noise factor for the source illumination is the shot noise:
\begin{equation}
    Noise_{TSR} = \sqrt{ S_{TSR} + D t + N_r^2 } \approx \sqrt{ S_{TSR} }
\end{equation}
While $t$ is the camera's total exposure time, $D$ represents the dark noise coefficient factor, and $N_r$ the read noise.
We then can say that the noise ratio between the signals:
\begin{equation}
    \frac{Noise_{TSR}}{Noise_T} = \sqrt{ \frac{ Signal_{TSR} } {Signal_{T} t + D t + N_r^2} } \leq \sqrt{ \frac{Signal_{TSR}}{S_T} }
\end{equation}

\textbf{The SNR ratio}
- According to the definition:
\begin{align}
    \frac{SNR_{TSR}}{SNR_T} = & \frac{ \frac{S_{TSR}}{Noise_{TSR}} }{ \frac{S_{T}}{Noise_{T}} } \geq
    \left( \frac{S_{TSR}}{S_T} \right)^{3/2} \geq \left[ \left(1+\alpha \right) \min\{\gamma_b, \gamma_g, \gamma_r \} \right]^{3/2} \nonumber
\end{align}
It means that the SNR improves significantly for increasing the $\alpha$ factor. For approximately white object and when $\alpha >> 1$ we get that improvement in the SNR goes like $\propto \alpha^{3/2}$.

\textbf{Dealing with environment illumination}
WhenWhen the environment illumination is not negligible ($\alpha \sim 1$), the previous assumption about the detected light does not hold anymore. Then, we can estimate the error in our method, saying that each channel has an additional detected light from the environment: 
\begin{equation}
    \vec{C} \Rightarrow \vec{C}+\Delta \vec{C} =\vec{C} + \frac{1}{\alpha}\vec{C}
\end{equation}
Which leads to a change in $\Delta I$:
\begin{equation}
    \vec{I} \Rightarrow \vec{I}+\Delta \vec{I}
\end{equation}
\begin{equation}
    \Delta \vec{I} = \frac{1}{\alpha} \textbf{M}^{-1} \textbf{S} \left( \textbf{S}^\intercal \textbf{M}^{-1} \textbf{S} \right)^{-1} \cdot \vec{C} = \alpha^{-1} \vec{I}
\end{equation}
So the error grows as $\alpha^{-1}$. It worth mentioning that the environment illumination contribution can be estimated before and then subtracted from the channel vectors.

\section{Numerical Simulations and Analysis}
In order to demonstrate our method, we built a computational simulator. The simulator simulates an ideal matt and white object, with no environment illumination, that performs any dynamics such that a particular pixel at the image can be described as a continuous trajectory of intensity versus the time. Apart from the scene, the simulator simulates the camera sampling method via integration and sampling in FPS and effective flickers in RGB colors. Everything is assumed to be ideal, such that in the presence of a red flicker (for example), there is no green and blue intensity value captured at all. Furthermore, we set the exposure time equal to one over the camera's frames-per-second, neglecting the sensor reading time delay (which is a good approximation for common cases in reality).
For each of the following results, unless otherwise mentioned, we simulated camera frame-per-second to be $10_{Hz}$ and we limit our analysis to the cases when $N=3$, $N=4$, $N=5$ and $N=6$ but it can be examined for higher up-sampling factor as well.


\subsection{Flicker pattern analysis}
The freedom to choose the flicker pattern still raises questions about which pattern should be chosen to maximize signal reconstruction performance. For example, for a specific channel (B, G, or R), one can choose whether to perform a flicker at one specific time step and then get as much information as possible about this specific time step (at that channel) or to apply the flicker for some time steps. Then the camera collects the accumulative values of this channel, which has some uncertainty about any specific time step. However, it gives information from a more extensive temporal range of the signal.
Two approaches have been examined here, the reconstruction error for randomly changing patterns over time and a comparison between some arbitrary flicker patterns.
For both analyses, we simulated 10,000 random sinus functions, with a temporal frequency of $[5_{Hz},30_{Hz}]$ (uniformly distributed) and a total duration of 5 seconds each.

\begin{figure}[h]
    \centering
    \includegraphics[trim={0cm 0cm 0cm 0cm},width=13.0cm, height=6.5cm]{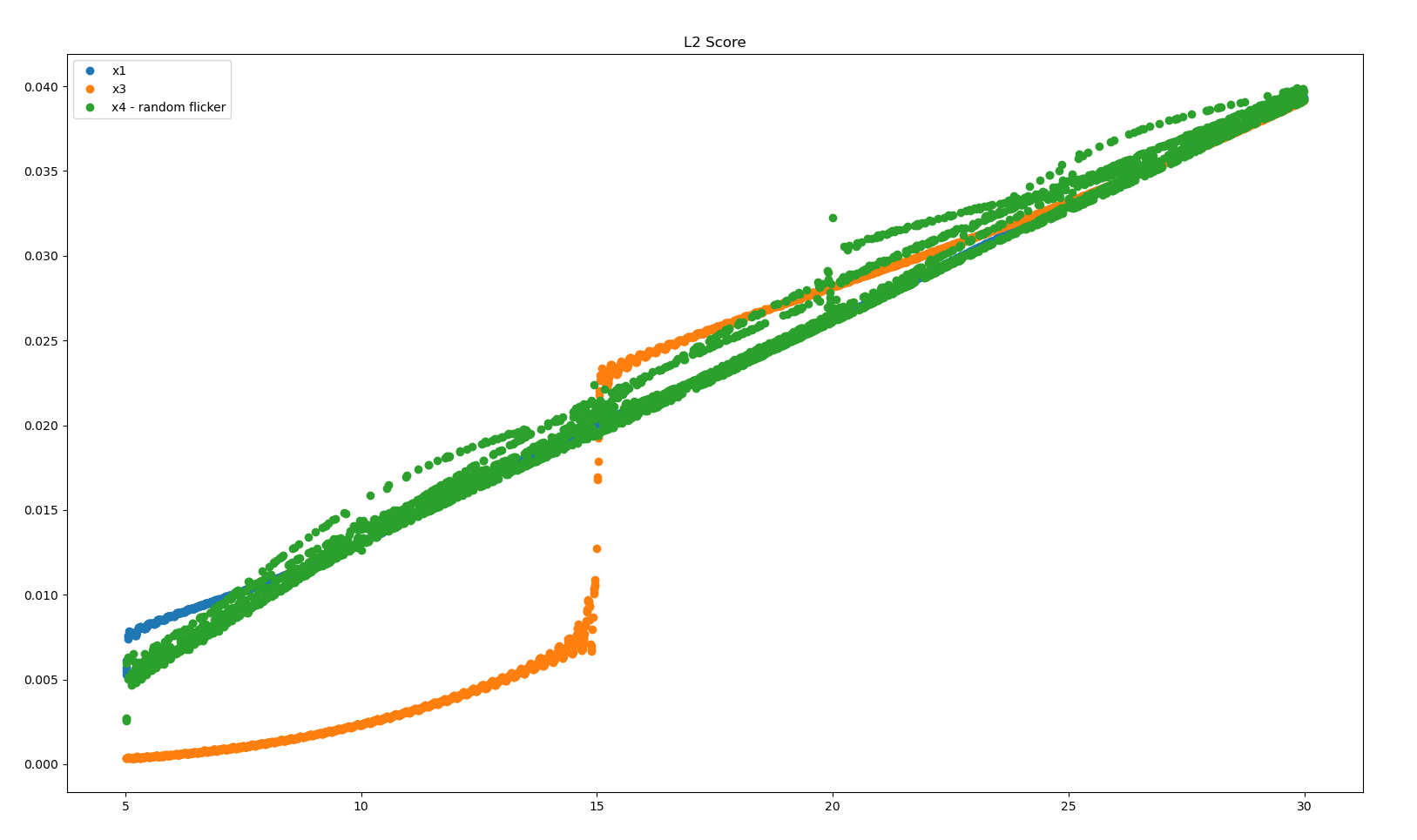}
    \caption{Graph of the L2 error vs. the frequency when using random S full-rank matrix for each frame. The result (in green) is represented compared to the non-up-sampled signal (blue) and compared to the particular case of $N=3$ reconstructed signal, and $S$ is the identity matrix (orange).}
    \label{fig:compare_random_matrix_4x}
\end{figure}

\textbf{Randomly Flicker} - This analysis has been done via random sampling of the flickering pattern. Practically, we sample full rank matrices $S$ for each frame and calculate the method L2 error, and the results are shown in figure \ref{fig:compare_random_matrix_4x}.
This random sampling was less effective since the error is close to the original camera sampled data's error. That is probably because changing the flicker pattern every frame does not enable the algorithm to detect high frequencies, requiring consistent small distance samples of the signal.

Our second test was to examine the reconstruction error for different fixed flicker patterns. Ideally, it is best to search among all the possible existing matrices, but this number is enormous, and we decided to focus on several specific flicker patterns for $N=4$, $N=5$ and $N=6$ (please see Appendix \ref{Appendix:Flicker_orders} to see the different choices).

\begin{figure}[h]
     \centering
         \centering
         \includegraphics[width=0.49\textwidth]{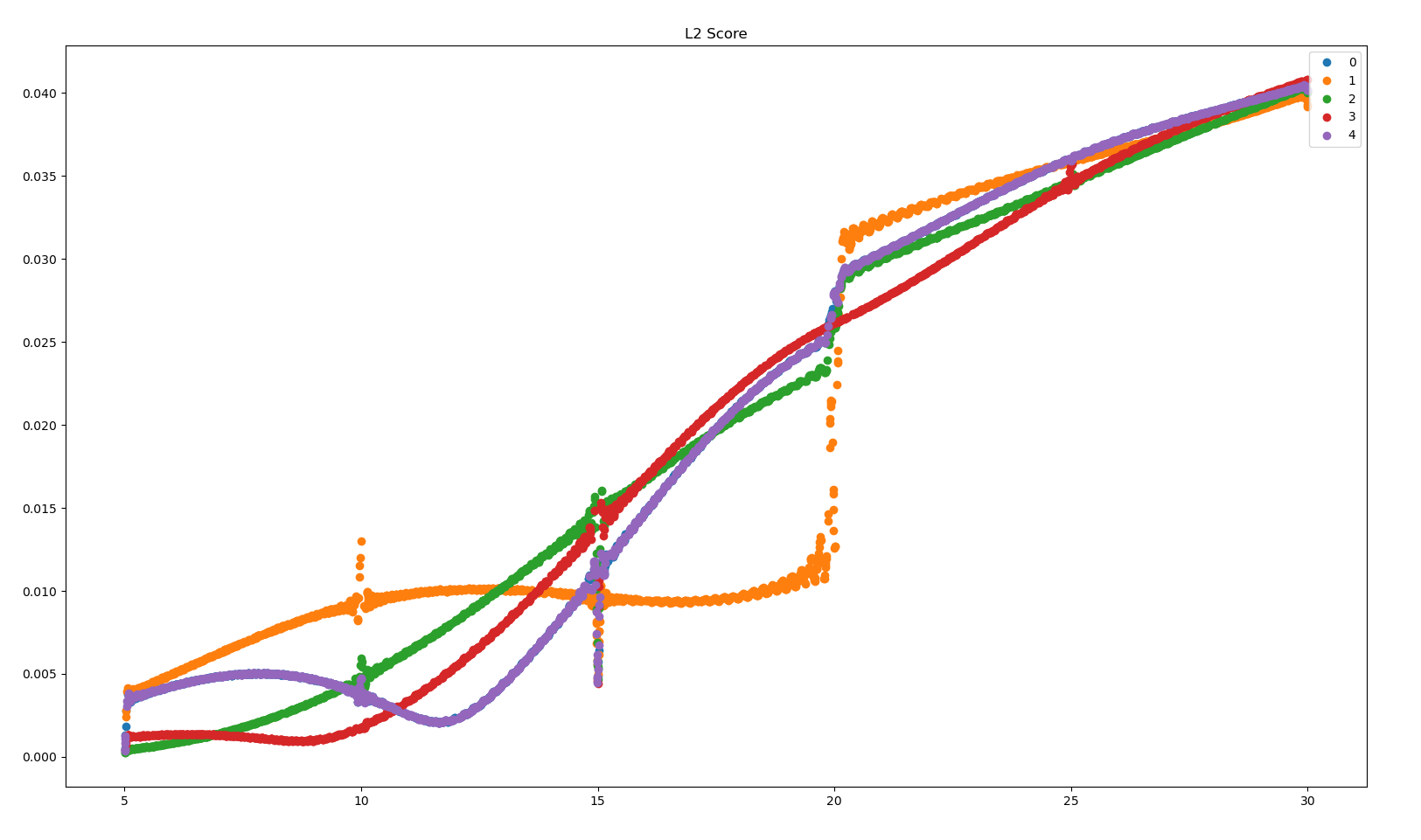}
     \hfill
         \centering
         \includegraphics[width=0.49\textwidth]{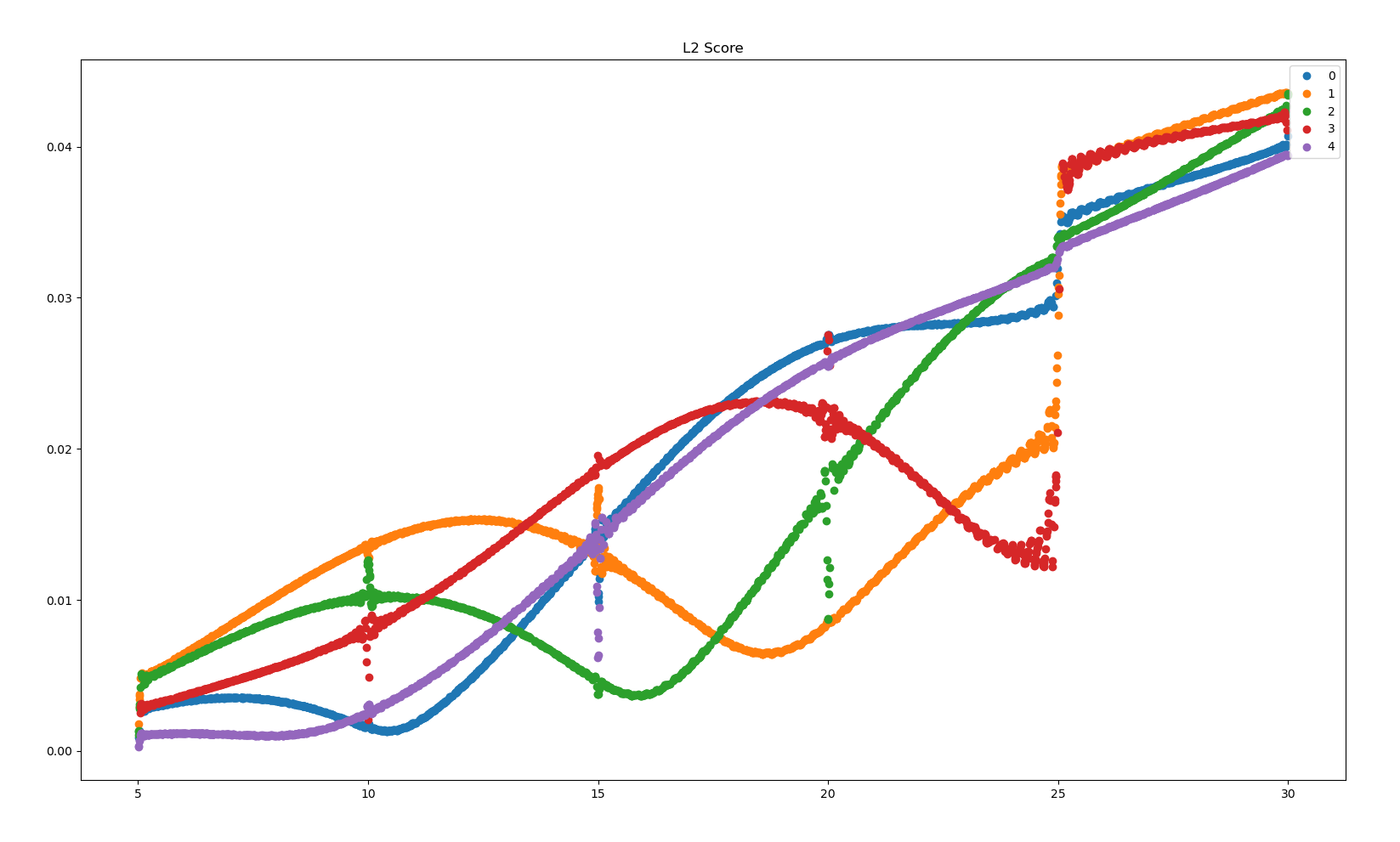}
     \hfill
         \centering
         \includegraphics[width=0.49\textwidth]{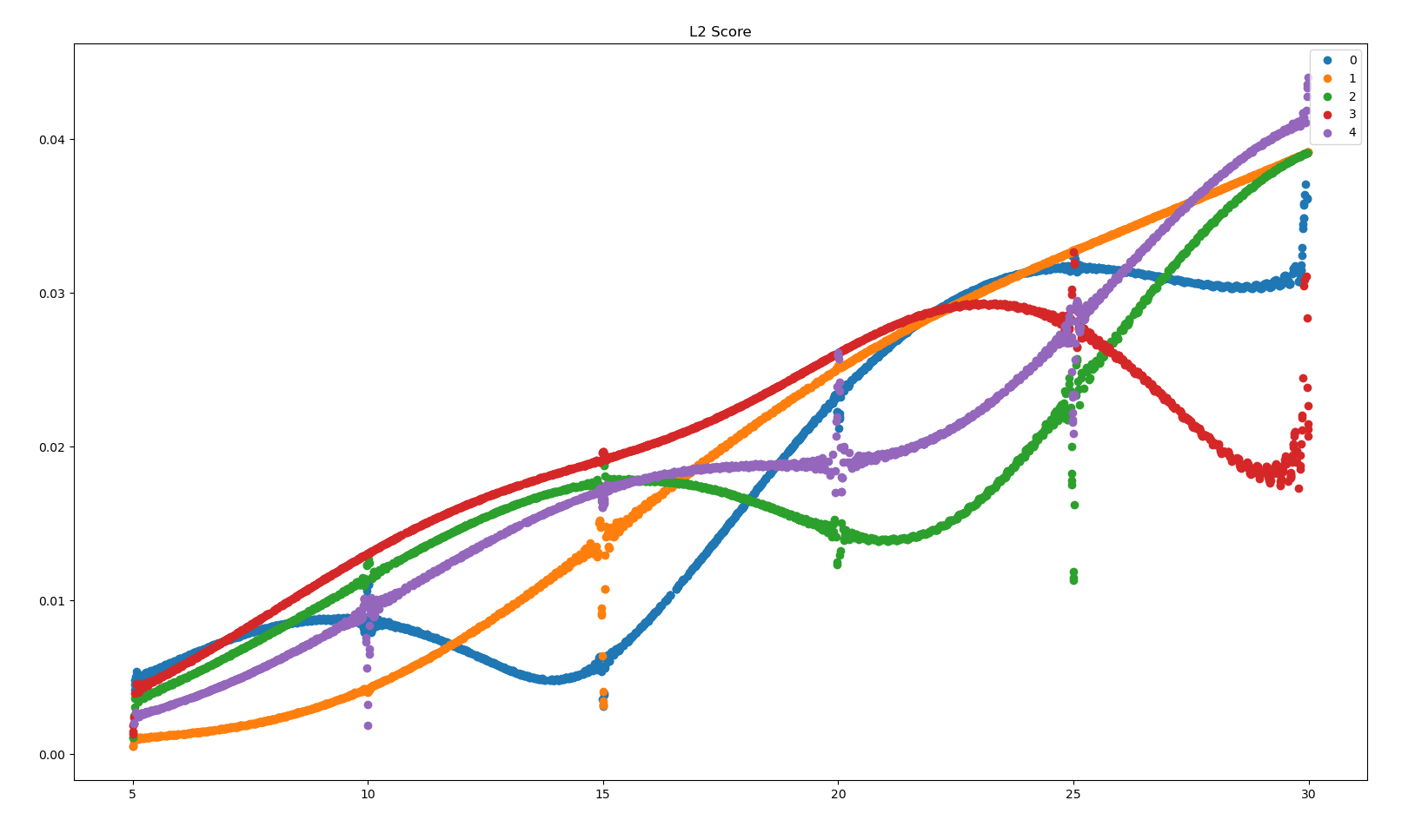}
    \caption{L2 error comparison between several candidates for flicker pattern, for different up-sample factor N (4, 5 and 6). Y-axis represents the error rate and X-axis represents the frequency.}
    \label{fig:compare_fixed_pattern}
\end{figure}

The results are represented in figure \ref{fig:compare_fixed_pattern}. For each N factor, the "jump" in error at a certain frequency ($20_{Hz}$,$25_{Hz}$,$30_{Hz}$ respectively) is due to the Nyquist theorem of sampling. For each $N$, there is no one specific graph that can be considered the best one among all candidates. Nevertheless, it is quite clear that if we focus on a specific spectral range, we can divide the spectrum into adjacent regions where each N gets its lowest error. For example: $N=3: [5_{Hz},15_{Hz}]$, $N=4: [15_{Hz},20_{Hz}]$, $N=5: [20_{Hz},25_{Hz}]$, $N=6: [25_{Hz},30_{Hz}]$, and due to that, we choose the best flicker pattern as follows: pattern 1 (for N = 4), pattern 3 (for N = 5) and pattern 4 (for N = 6). These results support our scanning method attitude for merging different temporal windows to construct the entire spectral domain.

An additional comparison is presented in the table \ref{tab:tsr_error_comp}.

\begin{table}[htbp]
\centering
\caption{Reconstruction error for each $N$ factor, for different frequencies ranges.}
\begin{tabular}{|| c | c | c | c ||}
\hline
\textbf{N} & \textbf{Frequencies} [Hz] & \textbf{L2 Error [$\times 10^{-3}$]} & \textbf{Normalized Error} \\
\hline
\hline
$3$ & $5-15$ & $\pm 3.1$ & $1$ \\
\hline
$4$ & $15-20$ & $\pm 10$ & $3.22$ \\
\hline
$5$ & $23-25$ & $\pm 14$ & $4.5$ \\
\hline
$6$ & $28-30$ & $\pm 18$ & $5.8$ \\
\hline
\end{tabular}
 \label{tab:tsr_error_comp}
\end{table}

\subsection{Simulations Results}
\begin{figure}[h]
    \centering
    \includegraphics[trim={0cm 0cm 0cm 0cm}, width=13.0cm, height=7cm]{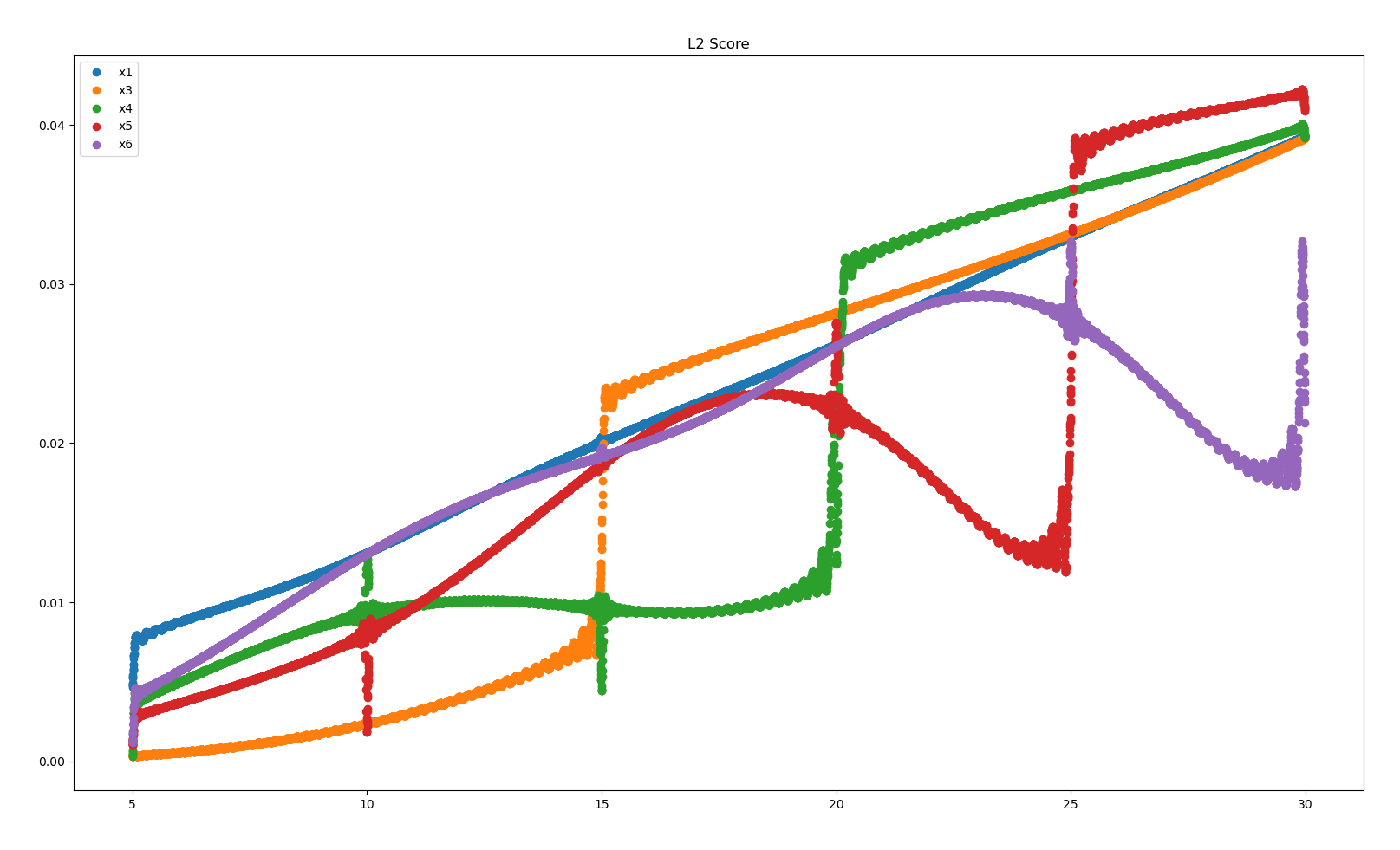}
    \caption{L2 error between the actual harmonic signal vs. different frequencies. Y-axis represents the error rate and X-axis represents the frequency. A comparison between various up-sampled factors. Frequency $5_{Hz}$ is the maximum the camera can detect due to Nyquist theorem, an up-sample factor of N = 3, 4, 5, 6 extends the frequency range to $15_{Hz}$, $20_{Hz}$, $25_{Hz}$ and $30_{Hz}$ respectively.}
    \label{fig:compareN_1}
\end{figure}

Here, we compare various $N$ factors with the signal reconstruction L2 error. The generated signal was, as in the previous section, via random sampling of 10,000 sinus functions at temporal frequencies of $[5_{Hz},30_{Hz}]$ (uniformly distributed).
The flicker patterns we chose to use, based on previous analysis, are:
\begin{align*}
N = 3: \hspace{0.1cm} &
\vec{b} = (1,0,0),
\vec{g} = (0,1,0),
\vec{r} = (0,0,1) \\
N = 4: \hspace{0.1cm} &
\vec{b} = (1,0,0,1),
\vec{g} = (1,0,1,0),
\vec{r} = (0,1,0,1) \\
N = 5: \hspace{0.1cm} &
\vec{b} = (0,1,0,0,0),
\vec{g} = (1,0,1,0,1),
\vec{r} = (0,0,0,1,0) \\
N = 6: \hspace{0.1cm} &
\vec{b} = (1,0,1,0,1,0),
\vec{g} = (0,1,0,1,0,1),
\vec{r} = (1,1,1,1,1,1)
\end{align*}
The results can be seen in figure \ref{fig:compareN_1}, where one can figure out several conclusions. First, the blue line (the linear curve) is the maximum error among all, and it is given by the camera's original signal with no up-sampling factor. Second, every up-sample factor extends the frequency detected range up to a different cut-off frequency due to the Nyquist sampling theorem. Third, the reconstruction quality for different N has a dependence on the frequency while each up-sampling factor reaches better results in different frequency regions. This insight might help a lot when there is some prior knowledge about the scene spectrum. Besides, these findings also support our presented scanning method technique.

\begin{figure}
     \centering
         \centering
         \includegraphics[width=\textwidth]{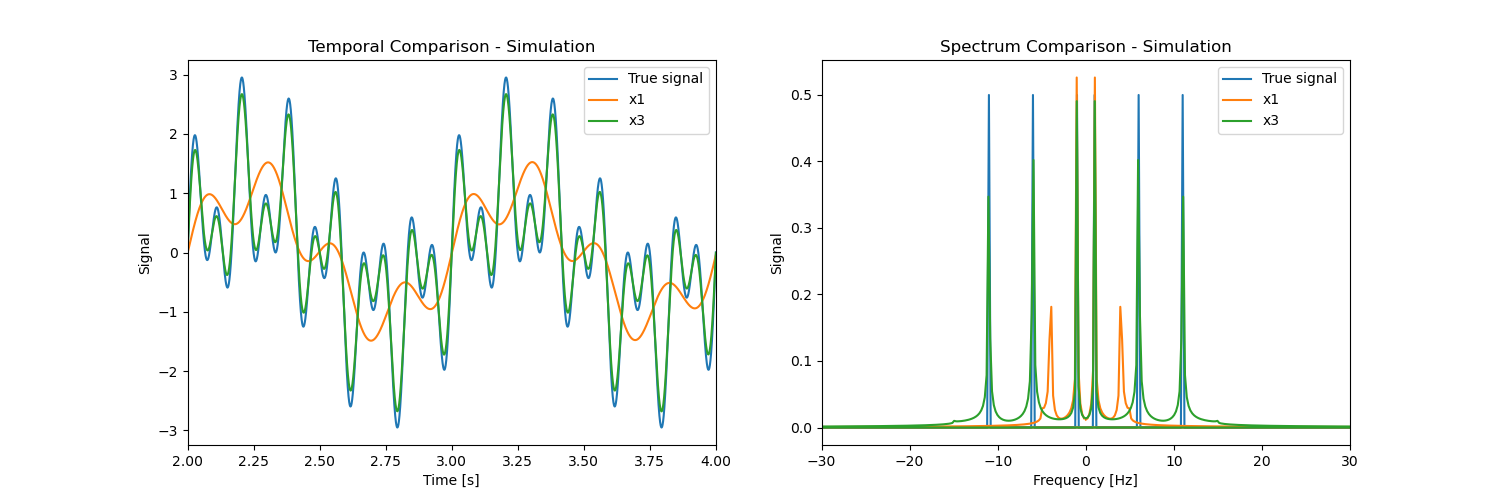}
     \hfill
         \centering
         \includegraphics[width=\textwidth]{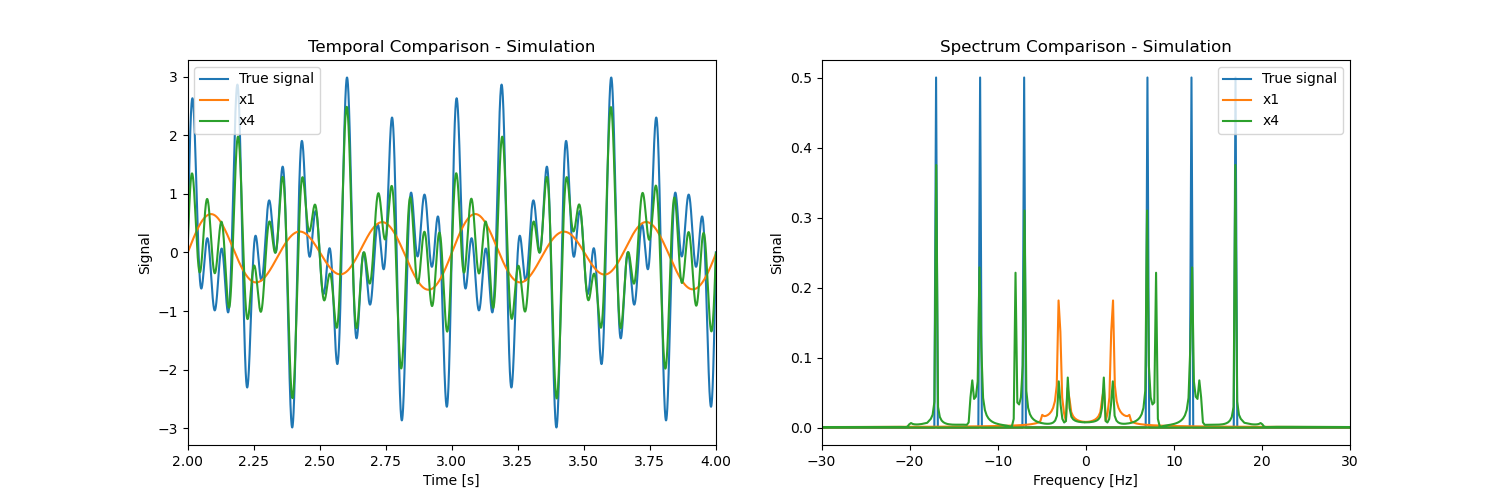}
     \hfill
         \centering
         \includegraphics[width=\textwidth]{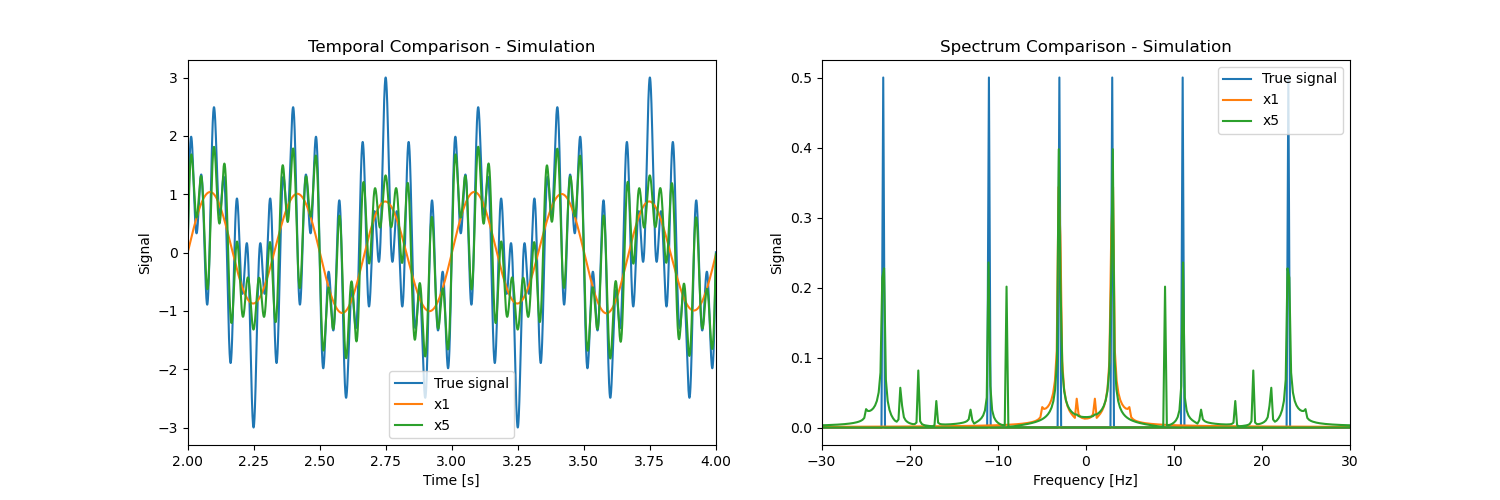}
     \hfill
         \centering
         \includegraphics[width=\textwidth]{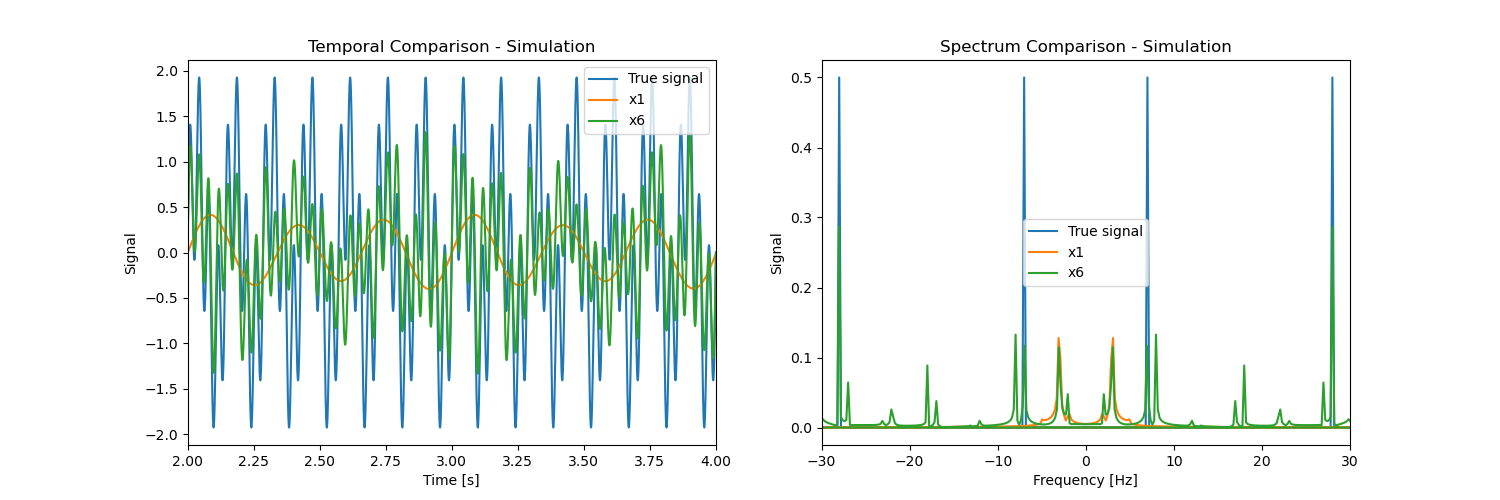}
    \caption{Simulation results. N = 3, 4, 5 and 6, camera FPS = 10. Blue - the original signal, Orange - camera reconstruction (no TSR), Green - our TSR algorithm.}
    \label{fig:simulation_N_3456}
\end{figure}

\textbf{Simulation results} - The simulation results are shown in figure \ref{fig:simulation_N_3456} when we simulated the following temporal signals:
\begin{align*}
N = 3: \hspace{0.1cm} & x(t) = \sin{(2 \pi t)}+0.3 \sin{(4 \pi t)}+0.8 \sin{(10 \pi t)} +0.25 \sin{(18 \pi t)}+0.75 \sin{(22 \pi t)}+0.5 \sin{(36 \pi t)} \\
N = 4: \hspace{0.1cm} & x(t) = \sin{(2 \pi t)}+\sin{(12 \pi t)}+\sin{(22 \pi t)} \\
N = 5: \hspace{0.1cm} &  
x(t) = \sin{(6 \pi t)}+\sin{(22 \pi t) }+\sin{(46 \pi t)}\\
N = 6: \hspace{0.1cm} & 
x(t) = \sin{(14 \pi t)}+\sin{(56 \pi t)}
\end{align*} 
And different flicker patterns:
\begin{align*}
N = 3: \hspace{0.1cm} &
\vec{b} = (1,0,0),
\vec{g} = (0,1,0),
\vec{r} = (0,0,1) \\
N = 4: \hspace{0.1cm} &
\vec{b} = (0,1,0,0),
\vec{g} = (1,0,0,1),
\vec{r} = (0,0,1,0) \\
N = 5: \hspace{0.1cm} &
\vec{b} = (0,1,0,0,0),
\vec{g} = (1,0,1,0,1),
\vec{r} = (0,0,0,1,0) \\
N = 6: \hspace{0.1cm} &
\vec{b} = (1,0,0,0,0,1),
\vec{g} = (0,1,1,0,0,0),
\vec{r} = (0,0,0,1,1,0)
\end{align*}

One can notice that our method significantly improves the ability to detect and reconstruct the signal spectral content even though one can still recognize the aliased signal parts.

To demonstrate this technique's performance and the Anti-Aliasing algorithm results, we simulated a signal 10 seconds long, and we defined 2, 3, and 4 temporal windows, each at size 5 seconds, 3.33 seconds, and 2.5 seconds respectively. Every temporal window had its own $N$ factor.

The simulated signal is:
\begin{equation}
    x(t) = SW_{12}(t)+SW_{19}(t)+SW_{23}(t)+SW_{27}(t)
\end{equation}
While we took the $N$ factors to be 3, 4, 5, and 6 (each corresponds to a temporal window). The result is shown at figure \ref{fig:scanning_method [3,4,5,6]}. To avoid white noise, we filtered out the lowest $5\%-10\%$ of the spectrum (filter uniform in the spectrum).

\begin{figure}
  \centering
  \includegraphics[width=1.0\linewidth]{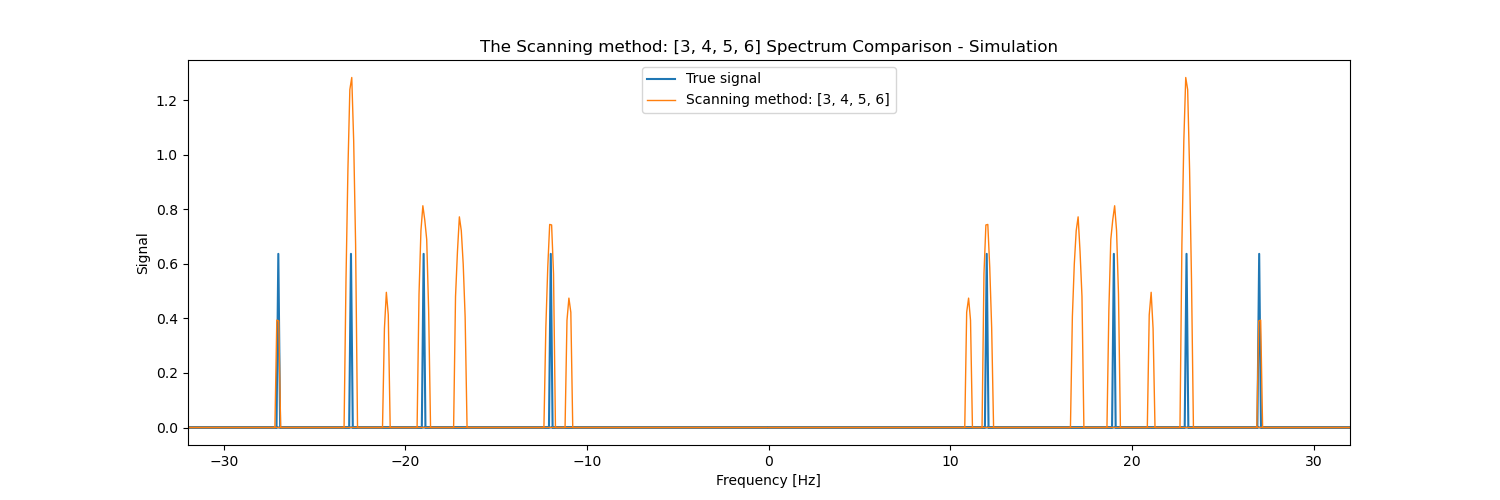}
  \centering
  \includegraphics[width=1.0\linewidth]{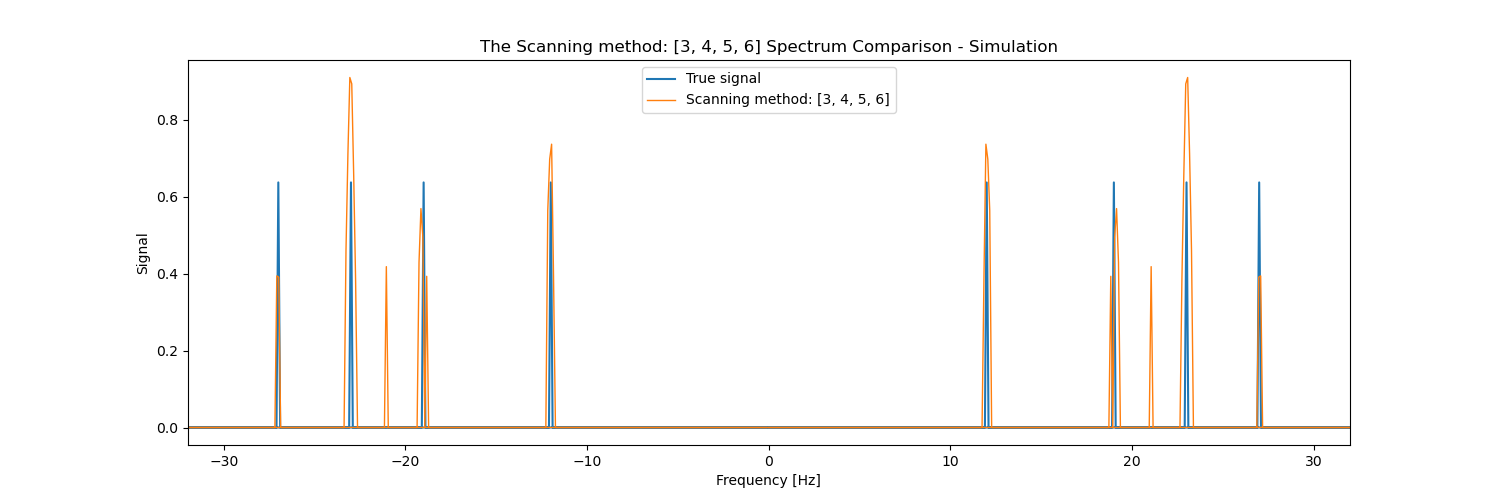}
\caption{Scanning Method, Top: a combination of N=3, N=4, N=5, and N=6 via scanning method, Bottom: after applying the Anti-Aliasing algorithm.}
\label{fig:scanning_method [3,4,5,6]}
\end{figure}

We can see that we get a good result using this technique and even additional improvement when using the Anti-Aliasing algorithm.

\section{Results and Discussions}

\begin{figure}
\centering
\includegraphics[width=0.6\textwidth]{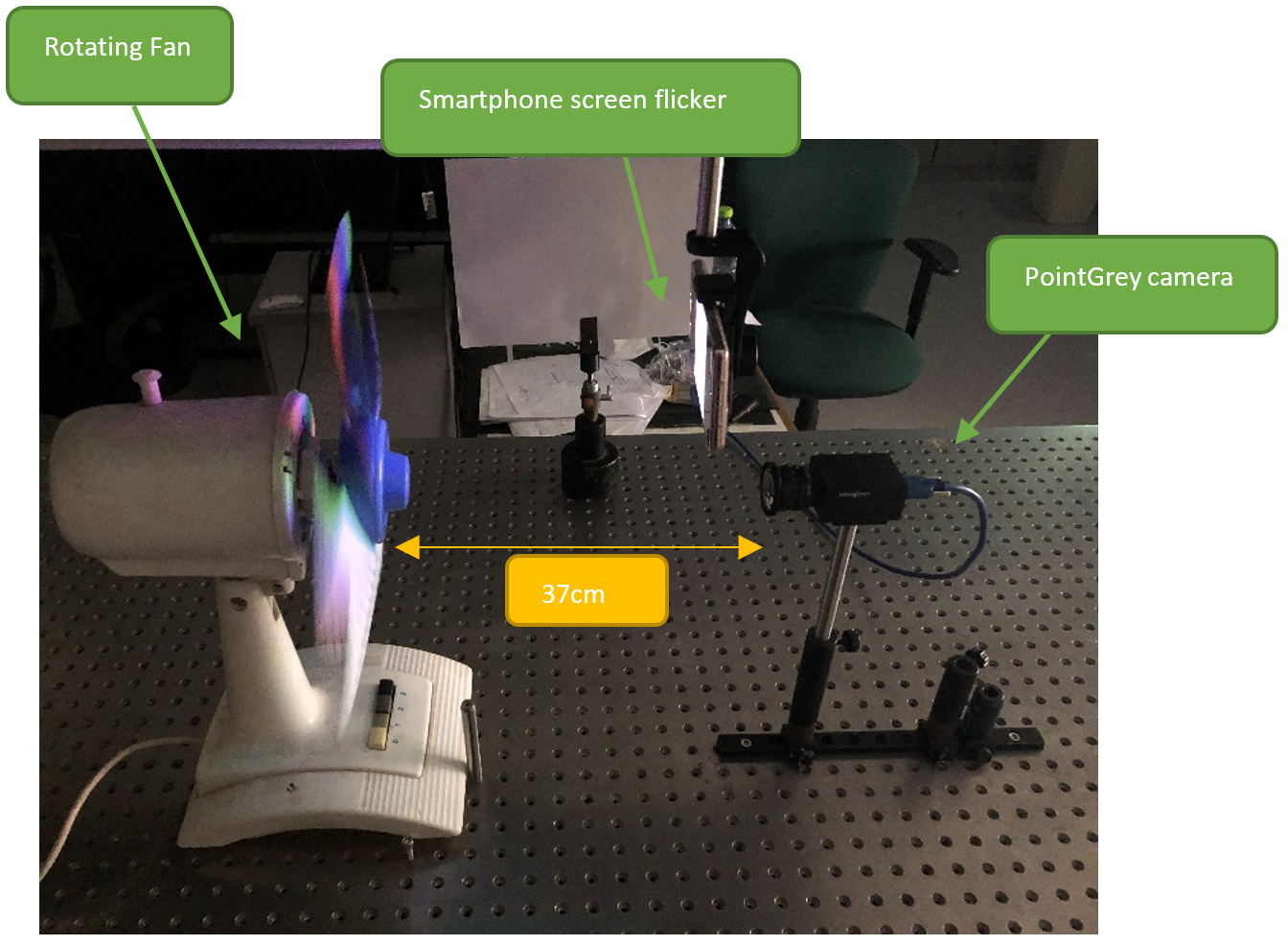}
\caption{Our experimental setup. We used a rotating fan (with a white paper sheet stuck to it), a smartphone screen as a colored flicker, and a camera.}
\label{fig:TSR_experimental_setup}
\end{figure}

Our experimental setup can be seen in figure \ref{fig:TSR_experimental_setup}. The recorded scene is a rotating fan at a constant frequency. A white paper sheet covered the fan blades, and we used a PointGrey camera, whose speed can be tuned. Our flicker is a smartphone screen Huawei P40, with a refresh rate of $60_{Hz}$, and we set the camera frame-per-second to $10_{Hz}$, $20_{Hz}$ or $80_{Hz}$. Since the scene is periodic, we compare our results between two measurements of the same scene with low-FPS recording and high-FPS recording. For every N, we used the same coded pattern as found to be best among the candidates presented in the simulations \ref{fig:compare_fixed_pattern}. We set the rotating fan frequency to approximately $\pm 21.5_{Hz}$. Also, we normalized the DC value for the different signals to focus only on the temporal variations. White noise filtering was applied for all the measured signals in this experiment to avoid noise artifacts. We set each spectrum amplitude below a given threshold to zero and set the threshold to be $5\%$ to $10\%$ for all the following measurements.

\textbf{Illumination correction} - since we compared the actual signal (which was evaluated in the high frame-per-second recording) with the same signal captured with a low frame-per-second (and up-sampled), a compensation gain for the high frame-per-second signal should be made to overcome the illumination difference due to the different exposure time. Another correction has been done, and it is due to the object color (the gamma-factors), representing the reflections for Red, Green, and Blue. To detect the gamma factors to balance the intensities for all colors, we used a reference measurement of a white target (the center of the fan) to calibrate the intensity values relative to it.

\subsection{Experimental Results}
\begin{figure}
     \centering
         \centering
         \includegraphics[width=\textwidth]{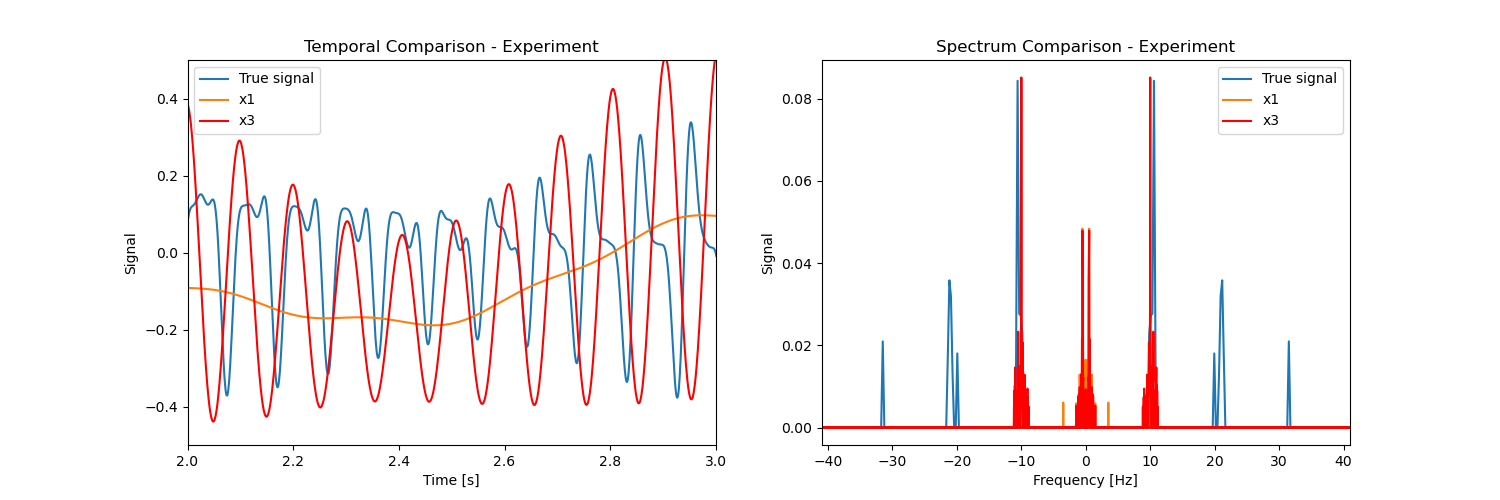}
     \hfill
         \centering
         \includegraphics[width=\textwidth]{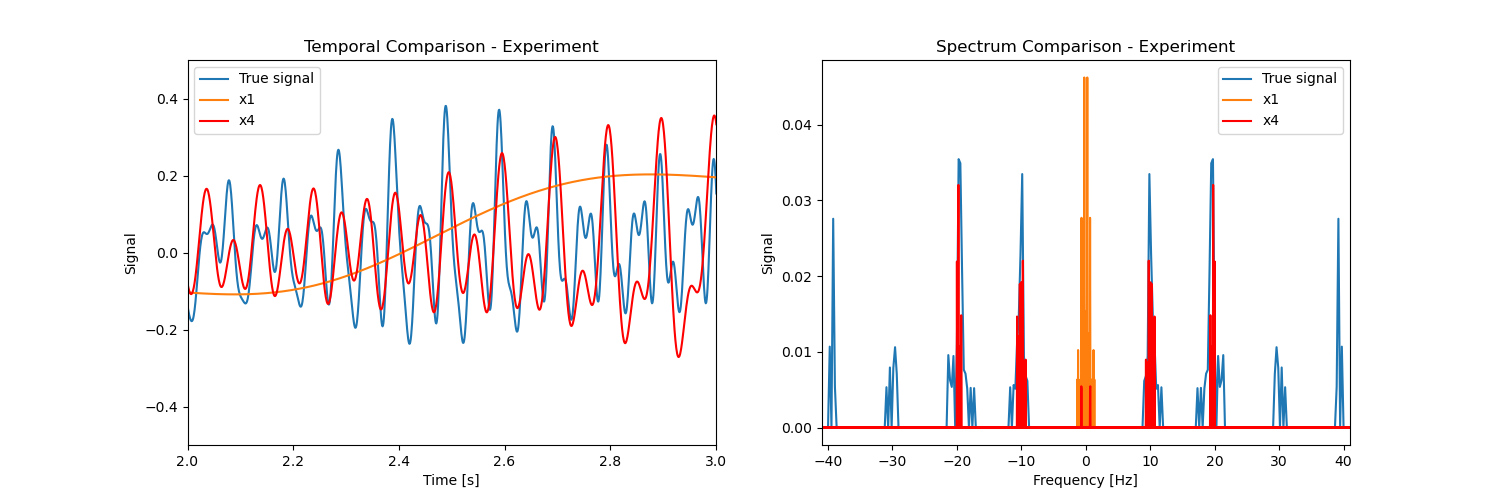}
     \hfill
         \centering
         \includegraphics[width=\textwidth]{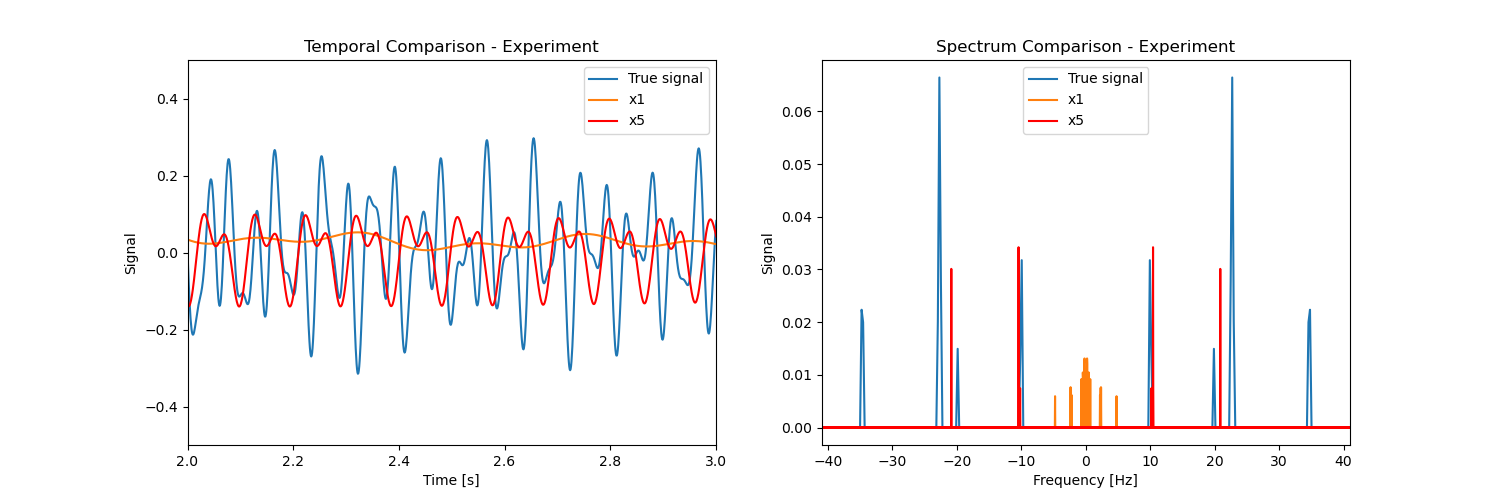}
     \hfill
         \centering
         \includegraphics[width=\textwidth]{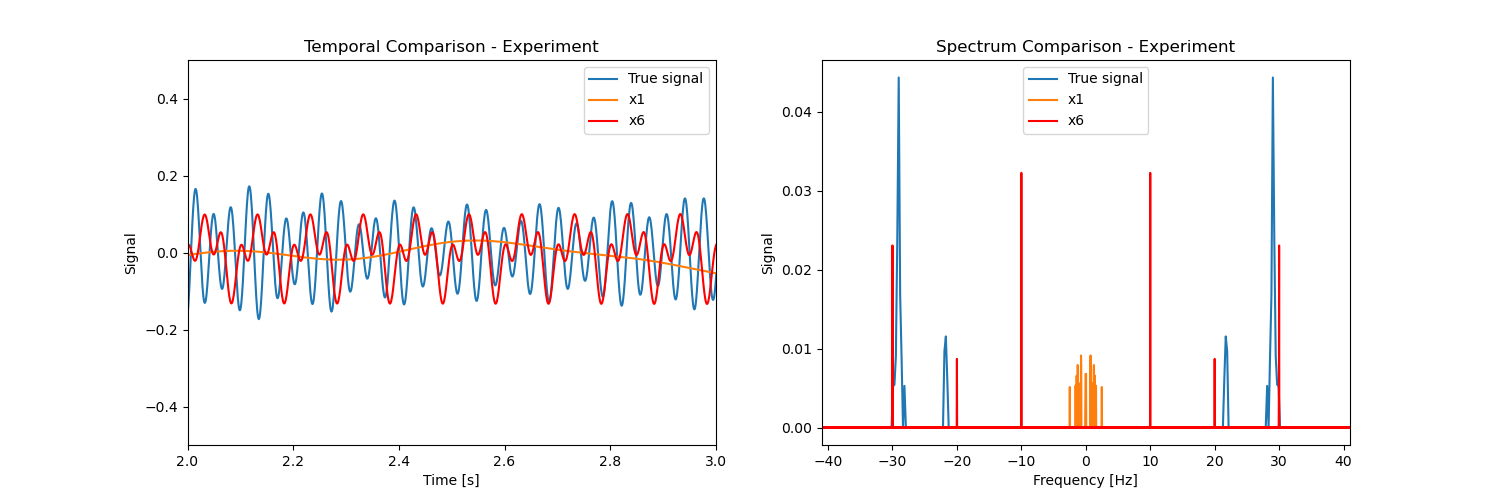}
    \caption{Experimental results. N = 3, 4, 5 and 6, camera FPS = 10. Blue - the original signal, Orange - camera reconstruction (no TSR), Red - our TSR algorithm. Our method successfully detects spectral components up to frequency $30_{Hz}$. }
    \label{fig:exp_N_3456}
\end{figure}

The experimental results is shown in figure \ref{fig:exp_N_3456} 

These results indicate that our method successfully detects high frequencies. Nevertheless, it can be seen from the graphs that sometimes there are some errors and artifacts in the result.

\subsection{Imaging results}
\begin{figure}
     \centering
         \centering
         \includegraphics [width=0.8\linewidth] {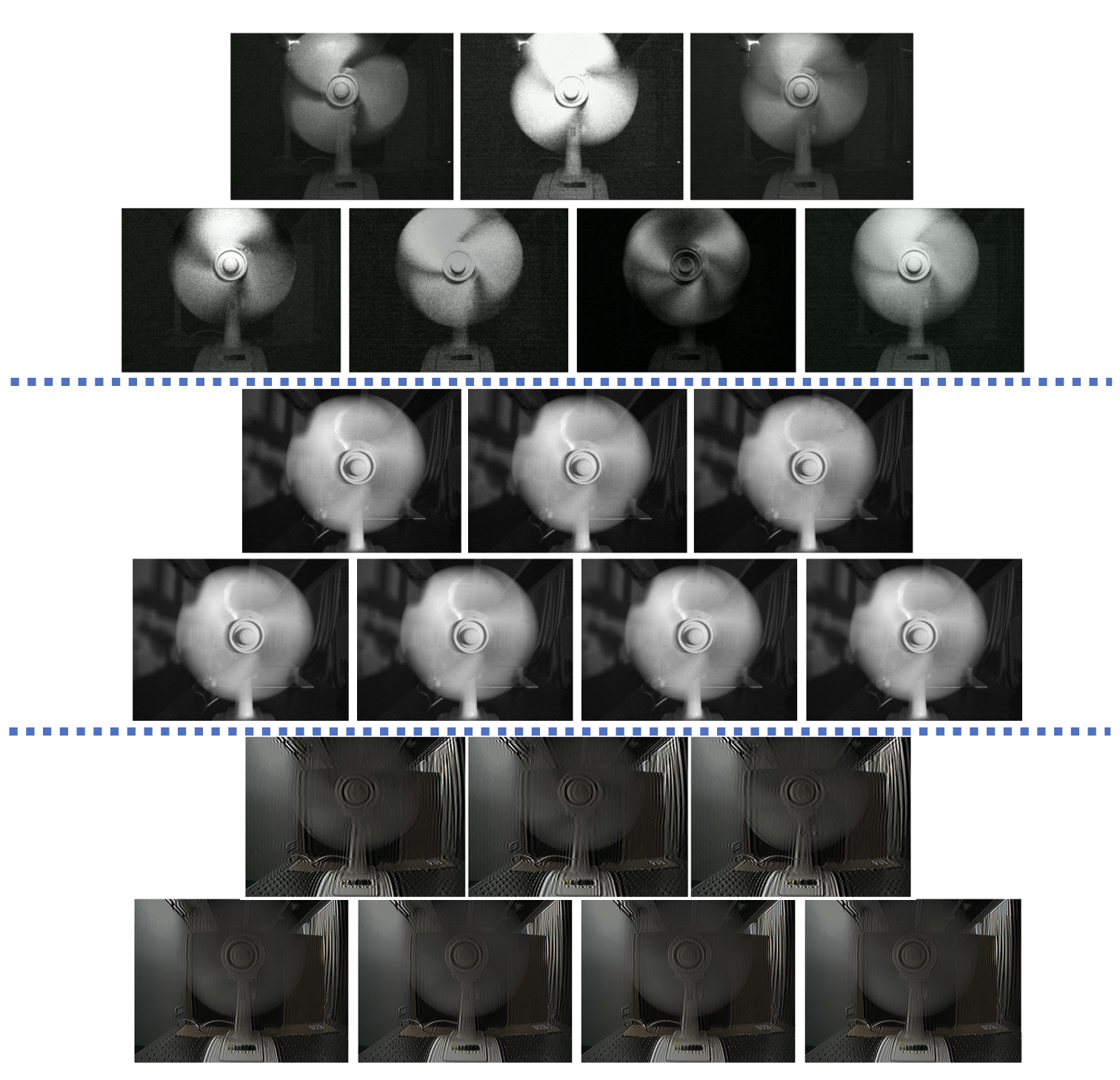}
    \caption{Different Imaging examples for N=3 and N=4. Top: Our technique, Middle: SuperSlowmo \cite{super_slowmo}, Bottom: Flatter Shutter \cite{tsr_debluring_shutter} }
    \label{fig:imaging_examples_tsr}
\end{figure}

Apart from the ability to capture high frequencies, we demonstrate in figure \ref{fig:imaging_examples_tsr} imaging results examples. Evidently, we found $W_{t}=3$ $W_{s}=1$ to be optimal. For comparison we used the SuperSlowmo algorithm \cite{super_slowmo} to raise the frame-per-second of the scene.

\subsection{SNR and performance result}
To evaluate the SNR for different $\alpha$ factors, we used a clean, white paper located $\pm 40_{cm}$ in front of the camera and the flicker. We used different environment illumination using a white-light projector and measured the illumination values using Lux-meter. The results are shown in figure \ref{fig:tsr_exp_snr}.
As one can notice, the SNR is improved since the flicker increases the light in the scene

An additional experiment was to measure the method reconstruction performance versus the $\alpha$ factor. Here we focused on $N=3$, and the results are shown in figure \ref{fig:tsr_snr_error_vs_alpha}.
Indeed, there is a degradation of the method performance when decreasing the $\alpha$ factor, which means increasing the illumination of the environment relative to the flicker illumination source.

\subsection{Motion estimation improvement}
\begin{figure}
\centering
\includegraphics[width=0.6\textwidth]{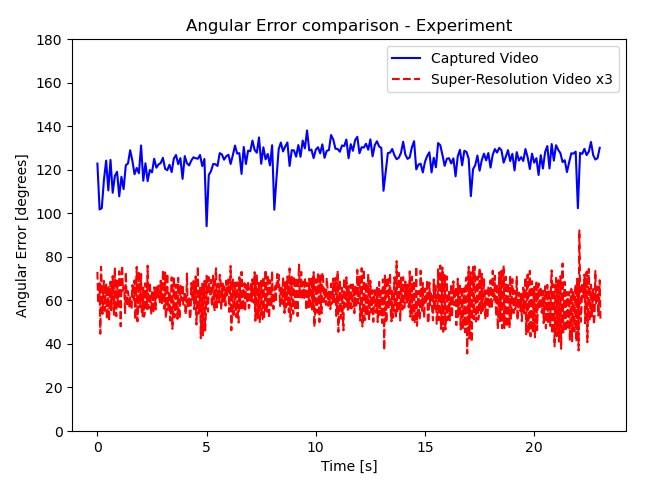}
\caption{The original video vs. the up-sampled version video ($N=3$) error comparison of motion estimation in time \cite{of_angular_error}. The lighting condition is poor, and the detection task is difficult, whereas still exist a significant improvement in the ability to detect the proper motion.}
\label{fig:upsampling_video_error_vs_time}
\end{figure}

One fundamental task in computer vision is motion estimation or optical flow estimation. Given the image's spatial and temporal derivatives, one can calculate the velocity of a pixel in the x-y plane. However, estimating the temporal derivative relies heavily on the camera frame-per-second rate.
We introduce here an application for our method. Since high temporal frequencies can not be detected in a low frame-per-second camera, applying our method and effectively raising the camera frame-per-second can improve the temporal. We measured the rotating fan's blade velocity (at the x-y plane) at each pixel and compared it to the ground truth, which was detected using a high frame-per-second camera. The result is shown in figure \ref{fig:upsampling_video_error_vs_time}

\subsection{Discussion}
\begin{figure}
    \centering
    \includegraphics[width=6.0cm]{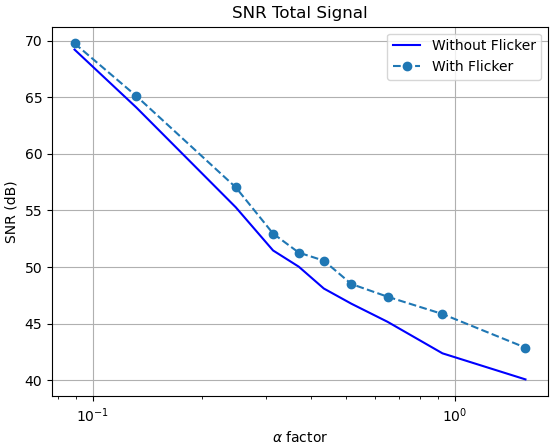}
    \qquad
    \includegraphics[width=6.0cm]{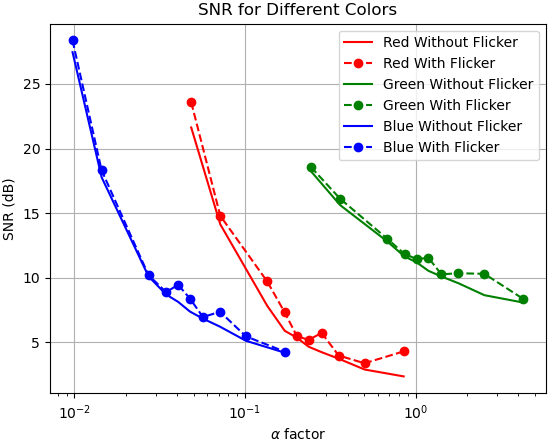}
    \caption{SNR measurement with and without flicker (vs. $\alpha$ factor). Please notice that $\alpha$ is on a logarithmic scale.}
    \label{fig:tsr_exp_snr}
\end{figure}

\begin{figure}
\centering
\includegraphics[width=0.6\textwidth]{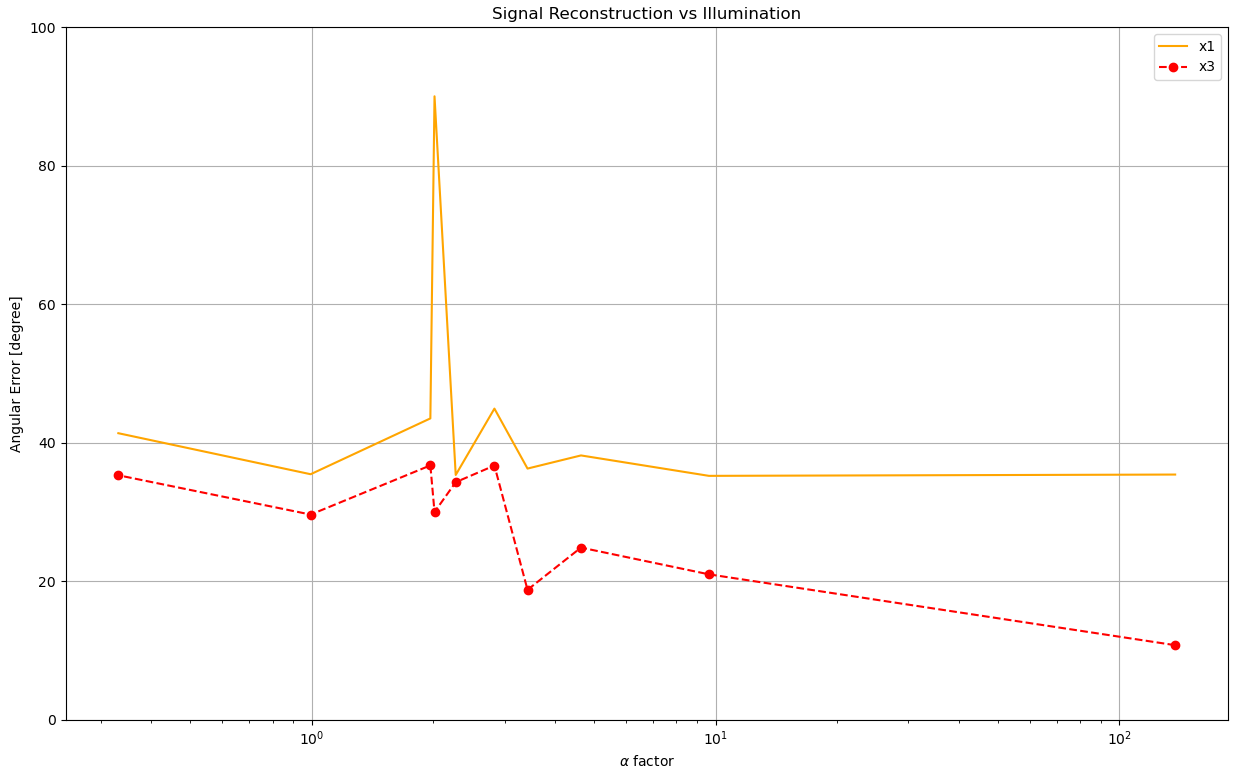}
\caption{Experimental measurements of cosine similarity between the actual signal and the reconstructed one for different $\alpha$ values note that $\alpha$ is in logarithmic scale ($N=3$).}
\label{fig:tsr_snr_error_vs_alpha}
\end{figure}

The results demonstrate how our method can significantly raise the temporal upper limit of the camera. We have reached the following conclusions:
\begin{itemize}
    \item The results demonstrate how our method can significantly raise the temporal upper limit of the camera.
    We found that applying different flickering patterns can deal with a significant change in the signal reconstruction error, and as expected, the greater the $N$ factor, the higher this error becomes.
    \item It has been shown how each flicker pattern can provide better accuracy at a particular frequency in the account of other frequencies.
    \item Following those findings, we introduced the scanning method, which has displayed good results, including aliasing attenuation.
    \item Our experiment showed, supported by theoretical derivation, how the SNR of the scene increases with our method. We demonstrated how our system performance improves when $\alpha$ factor improves \ref{fig:tsr_exp_snr}. Even though the error are still good even at low $\alpha$ \ref{fig:tsr_snr_error_vs_alpha}.
    \item We showed how motion estimation error decreases significantly when using our method \ref{fig:upsampling_video_error_vs_time}.
    \item From the experimental and the simulation results, it is clear that our method successfully detects high temporal frequencies. However, it still suffers from several issues and limitations which influence the reconstruction error, and we divide them into three:
    
    (1) Illumination correction issues - capturing the scene while the flicker mask is applied means that the scene is multiplied by the mask, and an illumination correction should be performed to extract the correct scene signal. We denoted this illumination correction as $\gamma_r$, $\gamma_g$, $\gamma_b$, and it depends on the flicker's color and intensity and the object's color. In our method, we assume we know the object's color so that those factors are known. Mismatch at those parameters might leads to some distortions in the signal. Practically, those parameters can be detected by one single scene reference image.
    
    (2) Temporal mismatch - When there is a mismatch in the synchronization between the flicker and the camera, then we get a shift in the flicker pattern relative to the camera exposure time. When this error occurs, we will likely see strong distortion in the signals, including a generation of new frequencies.
    
    (3) Reconstruction error - Our analysis has shown an inherent error factor in our method, especially for high $N$. This error component might lead to a new frequencies generation and signal distortions (as can be seen for $N=6$ at figure \ref{fig:exp_N_3456}).
    
    \item From our analysis, some patterns are better than others. One interesting question for future research is finding the system's optimal pattern and whether it depends on the scene frequency content.
\end{itemize}

\section{Conclusion}
In this work, we introduced a new method for Temporal super-resolution based on multi-channels flickering light sources. Our method demonstrated an excellent result in the simulation, while the accuracy of signal estimation was subjected to the $N$ up-sampling factor and the flicker patterns. We further demonstrated the performance of our scanning method and anti-aliasing technique. In our experiment, the results were good as well, and our method could extract very high frequencies (about a factor of 6) from the original camera Nyquist cut-off frequency. Also, we demonstrated experimentally how a motion estimation task is significantly improved thanks to our method.
While achieving temporal super-resolution is always accompanied by a trade-off between the accuracy results and the system complexity, here we demonstrated a method that constitutes a proper balance between the two. The up-sampling factor can go up to 6 (with moderate error) and even more, and the system complexity comes down to an additional colored flickering light source.

\bibliographystyle{plain}
\bibliography{references}

\section{Appendix}
\subsection{Solving the Optimization Problem}
In the general case we can say that:
\begin{equation}
    L = \sum_{n=1}^{N} \left( i_n-i_{n+1} \right)^2 + \sum_{m=1}^{M} \lambda_m \left( C_m - \sum_{n=1}^{N} i_n c_{m,n} \right)
\end{equation}
And for our case, we say that $B, G, R$ are the input digital values of colors Blue, Green and Red respectively. We shall define the following cost function:
\begin{gather*}
    L = \sum_{n=1}^{N} \left( i_n-i_{n+1} \right)^2  + \lambda_b \left( B - \sum_{n=1}^{N} i_n b_n \right) +
     \lambda_g \left( G - \sum_{n=1}^{N} i_n g_n \right) +
     \lambda_r \left( R - \sum_{n=1}^{N} i_n r_n \right)
\end{gather*}
While $\lambda_b, \lambda_g, \lambda_r$ are Lagrange Multipliers.
Derivative the Cost function with respect to the different parameters:
\begin{equation}
    \frac{\partial L}{\partial \lambda_b} = 0 \Rightarrow B = \sum_{n=1}^{N} i_n b_n
\end{equation}
\begin{equation}
    \frac{\partial L}{\partial \lambda_g} = 0 \Rightarrow G = \sum_{n=1}^{N} i_n g_n
\end{equation}
\begin{equation}
    \frac{\partial L}{\partial \lambda_r} = 0 \Rightarrow R = \sum_{n=1}^{N} i_n r_n
\end{equation}
$\forall 0 < k < N $:
\begin{gather*}
\frac{\partial L}{\partial i_k} = 2\left( i_k-i_{k+1} \right) - 2\left(i_{k-1} - i_k \right) - \lambda_b b_k - \lambda_g g_k - \lambda_r r_k = 0 \\
\Rightarrow 4i_k -2i_{k+1} - 2i_{k-1} = \lambda_b b_k + \lambda_g g_k + \lambda_r r_k
\end{gather*}
We would like to write these system of equations as vectors. Hence we define the following:
\begin{equation}
    \vec{ C }_{(3,1)} =
    \begin{pmatrix} 
        B \\ 
        G \\ 
        R 
    \end{pmatrix}
    \hspace{0.5cm}
    \vec{ \lambda }_{(3,1)} =
    \begin{pmatrix} 
        \lambda_b \\ 
        \lambda_g \\ 
        \lambda_r 
    \end{pmatrix}
    \hspace{0.5cm}
    \vec{ I }_{(N,1)} = \begin{pmatrix} 
    i_1 \\ 
    i_2 \\
    \vdots \\
    i_N 
    \end{pmatrix}
\end{equation}
\begin{equation}
    \textbf{ S }_{(N,3)} = \begin{pmatrix} 
     \vec{b} & \vec{g} & \vec{r} \\ 
    \end{pmatrix}
    \hspace{0.25cm}
    \textbf{ M }_{(N,N)} = \begin{pmatrix} 
    4 & -2 & 0  &\dots & 0\\ 
    -2 & 4 & -2 &\dots & 0\\ 
    0 & -2 & 4 & \dots & 0\\ 
    & & \vdots & & & \\
    0 & \dots & -2 & 4 & -2 \\ 
    0 & \dots & 0 & -2 & 4 \\ 
    \end{pmatrix}
\end{equation}
\begin{equation}
    B = \vec{b}^\intercal \cdot \vec{I}
    \hspace{1cm}
    G = \vec{g}^\intercal \cdot \vec{I}
    \hspace{1cm}
    R = \vec{r}^\intercal \cdot \vec{I}
    \hspace{1cm}
    \textbf{M} \cdot \vec{I} = \textbf{S} \cdot \vec{\lambda}
    \hspace{1cm}
\end{equation}
Notice that $\textbf{M}$ is a full-rank matrix (adding every row half of the row above leads to an upper Triangular matrix), so we can say:
\begin{align*}\label{eq:pareto mle2}
\vec{I} =\textbf{M}^{-1} \textbf{S} \cdot \vec{\lambda} \Rightarrow \hspace{1cm}& 
B = \vec{b}^\intercal \cdot \textbf{M}^{-1} \textbf{S} \cdot\vec{\lambda} \\
& G = \vec{g}^\intercal \cdot \textbf{M}^{-1} \textbf{S} \cdot\vec{\lambda} \\
& R = \vec{r}^\intercal \cdot \textbf{M}^{-1} \textbf{S} \cdot\vec{\lambda} \\
\end{align*}
And we can write it as:
\begin{equation}
    \vec{C} = \textbf{S}^\intercal \textbf{M}^{-1} \textbf{S} \cdot\vec{\lambda}
\end{equation}
Requiring that $S$ will be a full rank matrix and inverting the relation to find Lagrange multipliers:
\begin{equation}
    \vec{\lambda} = \left( \textbf{S}^\intercal \textbf{M}^{-1} \textbf{S} \right)^{-1} \cdot\vec{C}
\end{equation}
And the final result:
\begin{equation}
    \boxed{ \vec{I} = \textbf{M}^{-1} \textbf{S} \left( \textbf{S}^\intercal \textbf{M}^{-1} \textbf{S} \right)^{-1} \cdot\vec{C} }
\end{equation}
This compact results provides us a way, given a flicker vectors $\vec{b}, \vec{g}, \vec{r}$, to extract the signal values into each frame $\vec{I}$ using the total input values of Blue, Green and Red intensity.
Note that we have required that $S$ matrix, which represents the flicker pattern, will be a full rank matrix. This requirement limits the number of possible matrices, assuming $S_{ij}$ can only be 0 or 1. In general, it is not obvious what is $S$ matrix should be under these constrains because it very much depends on the sampled function behaviour manner. We will present later an analysis regarding choosing the flicker pattern.

\subsection{Expanding to spatial regularization}
To enhance the spatial correlation between adjacent pixels the cost function can be modified to add some spatial regularization. We assume that only the first-level neighbors are relevant and we define the domain P to be the 5 pixels domain (see figure \ref{fig:pixels_neighbors} )

\begin{gather*}
    \mathcal{L} = \sum_{n=1}^{N-1} \sum_{x,y \in P} w^{t}_{x,y} \left( i_{x,y,n}-i_{x,y,n+1} \right)^2 + 
    w^{s}_{x,y} \left( i_{x,y,n}-i_{x,y+1,n} \right)^2 + \\
    w^{s}_{x,y} \left( i_{x,y,n}-i_{x+1,y,n} \right)^2 +  \sum_{m=1}^{M} \sum_{x,y \in Patch} \lambda_{m,x,y} \left( C_{m,x,y} - \sum_{n=1}^{N} i_{x,y,n} c_{x,y,m,n} \right)
\end{gather*}
While $w^t_{x,y}, w^s_{x,y}$ are the weight factors for the temporal condition and the spatial condition, respectively. For simplicity we assume constant weight $w^s_{x,y}=w^s, w^t_{x,y} = w^t$.
This system has the same solution form under mapping the different vectors to be column stack vectors (for each of the pixels) for example:
\begin{equation}
    \vec{ C }_{(3*5,1)} =
    \begin{pmatrix} 
        B_{1} \\ 
        G_{1} \\ 
        R_{1} \\
        \vdots \\
        B_{5} \\ 
        G_{5} \\ 
        R_{5} \\
    \end{pmatrix}
    \hspace{0.5cm}
    \vec{ \lambda }_{(3*5,1)} =
    \begin{pmatrix} 
        \lambda_{b,1} \\ 
        \lambda_{g,1} \\ 
        \lambda_{r,1} \\
        \vdots \\
        \lambda_{b,5} \\ 
        \lambda_{g,5} \\ 
        \lambda_{r,5} \\
    \end{pmatrix}
\end{equation}
S matrix ($5Nx15$) is changed to be a block diagonal matrix while each block correspond to different pixel in the domain P.
M matrix ($5Nx5N$) has changed to be:
\begin{equation}
    \textbf{ M }_{i,j} = 
    \begin{cases}
        2w^s+2w^t & \text{if} \ i = j  \\
        -2w^t & \text{if} \ |i - j| = 1  \\
        -2w^s& \text{if} \ |i - j| mod 5 = 0  \\
        0 & \textbf{else} \ \\
    \end{cases}
\end{equation}

\subsection{Flickers order choices}
\label{Appendix:Flicker_orders}

\underline{N = 4}:
\begin{align*}
0: \hspace{0.1cm} &
\Vec{b} = (1,0,0,1), 
\Vec{g} = (0,1,0,0), 
\Vec{r} = (0,0,1,0) \\
1: \hspace{0.1cm} &
\Vec{b} = (1,0,0,1), 
\Vec{g} = (1,0,1,0), 
\Vec{r} = (0,1,0,1) \\
2: \hspace{0.1cm} &
\Vec{b} = (1,0,0,0), 
\Vec{g} = (0,1,1,0), 
\Vec{r} = (0,0,0,1) \\
3: \hspace{0.1cm} &
\Vec{b} = (1,1,0,0), 
\Vec{g} = (0,1,1,0), 
\Vec{r} = (0,0,1,1) \\
4: \hspace{0.1cm} &
\Vec{b} = (0,0,1,0), 
\Vec{g} = (0,1,0,0), 
\Vec{r} = (1,1,1,1) \\
\end{align*}
\underline{N = 5}:
\begin{align*}
0: \hspace{0.1cm} &
\Vec{b} = (1, 0, 0, 0, 1), 
\Vec{g} = (0, 1, 1, 0, 0), 
\Vec{r} = (0, 0, 1, 1, 0) \\
1: \hspace{0.1cm} &
\Vec{b} = (1, 0, 0, 1, 0), 
\Vec{g} = (1, 0, 1, 0, 1), 
\Vec{r} = (0, 1, 0, 0, 1) \\
2: \hspace{0.1cm} &
\Vec{b} = (1, 0, 0, 1, 0), 
\Vec{g} = (0, 0, 1, 0, 0), 
\Vec{r} = (0, 1, 0, 0, 1) \\
3: \hspace{0.1cm} &
\Vec{b} = (0, 1, 0, 0, 0), 
\Vec{g} = (1, 0, 1, 0, 1), 
\Vec{r} = (0, 0, 0, 1, 0) \\
4: \hspace{0.1cm} &
\Vec{b} = (1, 1, 0, 0, 0), 
\Vec{g} = (0, 1, 1, 1, 0), 
\Vec{r} = (0, 0, 0, 1, 1) \\
\end{align*}
\underline{N = 6}:
\begin{align*}
0: \hspace{0.1cm} &
\Vec{b} = (1, 0, 0, 0, 1, 0), 
\Vec{g} = (0, 1, 0, 0, 0, 1), 
\Vec{r} = (0, 0, 1, 1, 0, 0) \\
1: \hspace{0.1cm} &
\Vec{b} = (1, 1, 0, 0, 0, 0), 
\Vec{g} = (0, 0, 1, 1, 0, 0), 
\Vec{r} = (0, 0, 0, 0, 1, 1) \\
2: \hspace{0.1cm} &
\Vec{b} = (1, 0, 0, 1, 0, 0), 
\Vec{g} = (0, 1, 1, 1, 1, 0), 
\Vec{r} = (0, 0, 1, 0, 0, 1) \\
3: \hspace{0.1cm} &
\Vec{b} = (1, 0, 1, 0, 1, 0), 
\Vec{g} = (0, 1, 0, 1, 0, 1), 
\Vec{r} = (1, 1, 1, 1, 1, 1) \\
4: \hspace{0.1cm} &
\Vec{b} = (0, 1, 0, 0, 0, 0), 
\Vec{g} = (1, 0, 1, 1, 0, 1), 
\Vec{r} = (0, 0, 0, 0, 1, 0) \\
\end{align*}

\begin{figure} [h]
\centering
\includegraphics[width=1.0\linewidth]{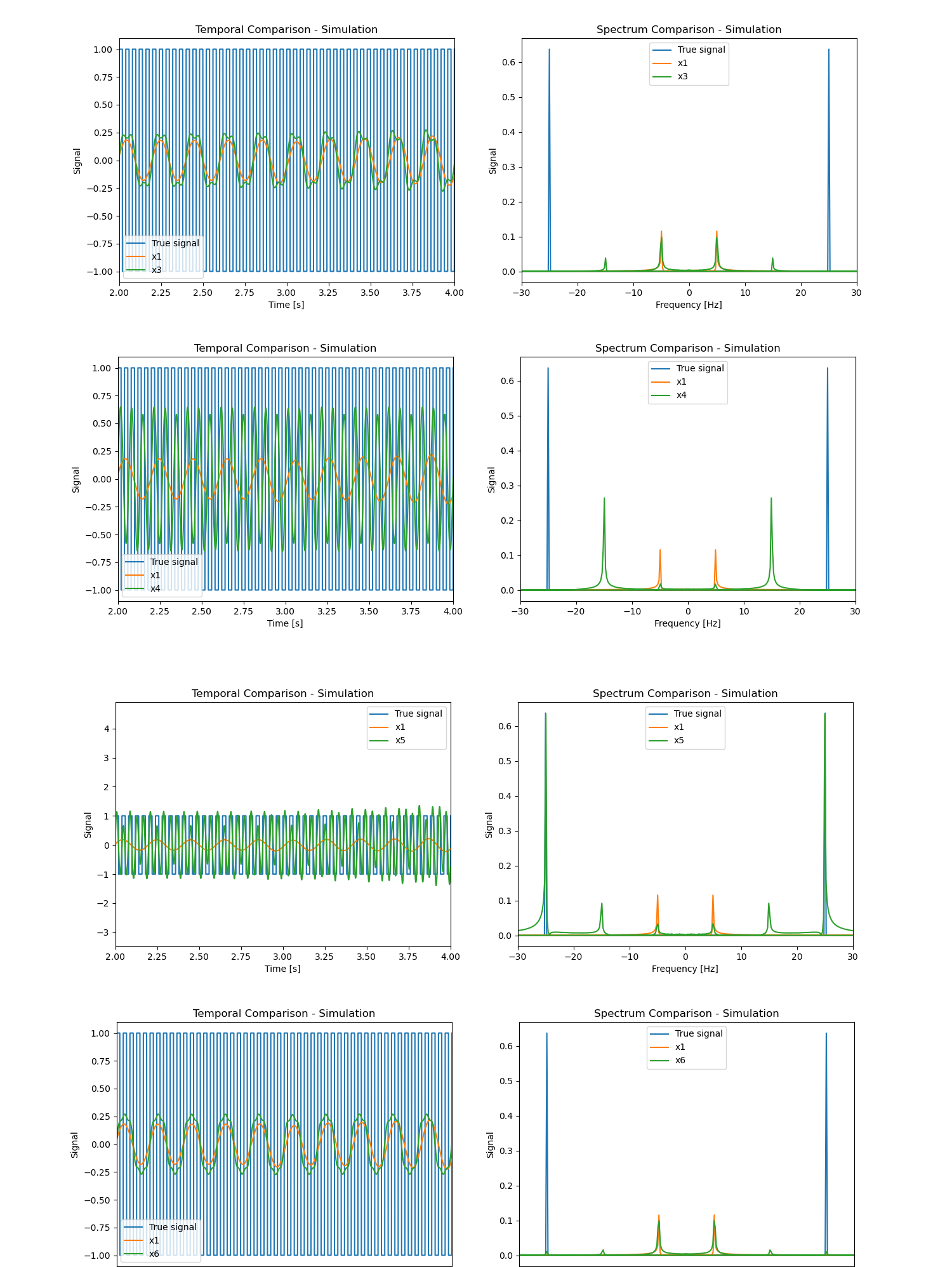}
\caption{Result comparison for square wave of 25 Hz}
\end{figure}

\begin{figure} [h]
\centering
\includegraphics[width=1.0\linewidth]{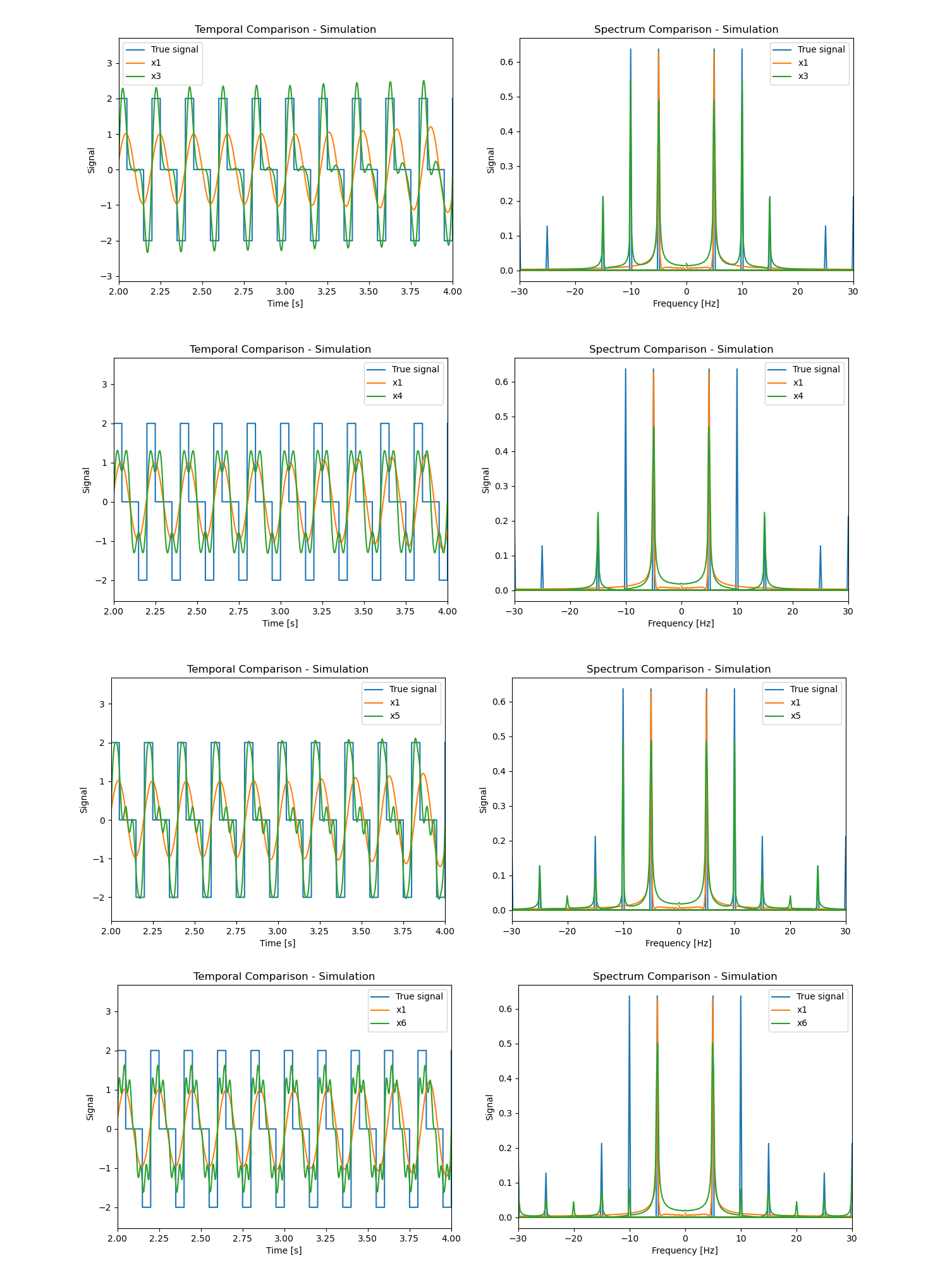}
\caption{Result comparison for square wave of 5 Hz and 10 Hz}
\end{figure}


\end{document}